# Integration of a Phosphatase Cascade with the MAP Kinase Pathway Provides for a Novel Signal Processing Function

Virendra K. Chaudhri[1], Dhiraj Kumar[1], Manjari Misra[1], Raina Dua[1] and Kanury V.S. Rao[1,2]

[1]Immunology Group
International Centre for Genetic Engineering and Biotechnology
Aruna Asaf Ali Marg
New Delhi – 110067
INDIA

We mathematically modeled the receptor-activated MAP kinase signaling by incorporating the regulation through cellular phosphatases. Activation induced the alignment of a phosphatase cascade in parallel with the MAP kinase pathway. A novel regulatory motif was thus generated, providing for the combinatorial control of each MAPK intermediate. This ensured a non-linear mode of signal transmission with the output being shaped by the balance between the strength of input signal, and the activity gradient along the phosphatase axis. Shifts in this balance yielded modulations in topology of the motif, thereby expanding the repertoire of output responses. Thus we identify an added dimension to signal processing, wherein the output response to an external stimulus is additionally filtered through indicators that define the phenotypic status of the cell.

[2]To whom correspondence should be addressed.
Email: kanury@icgeb.res.in
Tel: 0091-11-26741680. FAX: 0091-11-26715114

**INTRODUCTION.**

The prototypic Mitogen Activated Protein Kinase (MAPK) pathway consists of a three-tiered module composed of a MAPK (ERK-1/2), which is activated via phosphorylation by a MAPKK (MEK-1/2), which in turn is phosphorylated by a MAPKKK (Raf). This pathway exists in all eukaryotic organisms and controls fundamental cellular processes such as proliferation, differentiation, migration, survival, and apoptosis [1-3]. Importantly, the cascade arrangement of this module permits integration of a wide range of conserved cellular process, thereby enabling a precise control of the amplitude, kinetics, and duration of ERK-1/2 (ERK) activation. Several studies have documented that it is the ability to modulate these individual parameters of ERK activation that confers signaling specificity to this pathway, in terms of regulating the cellular response [4-7].

The contrast between the apparent biochemical simplicity of this pathway, and the range of complex cellular functions that it controls, has prompted extensive investigations on the structure and regulation of this pathway. One approach towards this goal has been through computational modeling of this pathway. The earliest model by Huang and Ferrell demonstrated that ERK activation exhibited ultrasensitivity, and that this was primarily due to the distributive mechanism involved in the dual phosphorylation of ERK [8,9]. Subsequent to these pioneering findings, an increasing number of models that have gained both in size and complexity have been developed over the years. The various aspects examined by these models include feedback cooperativity and inhibition, hysteresis, oscillations, Ras activation, scaffolding proteins, signal specificity and robustness among others [10-13]. From the standpoint of input/output relationships, these models have provided important new insights on feedback, sequestration, and scaffolding influences on the ERK response [14-17].

A less explored aspect of MAPK regulation, however, has been the role of phosphatases in regulating signal specificity [18]. An earlier study by Bhalla and co-workers iden-tified MAPK phosphatase (MKP) as the locus of flexibility that regulated between monostable and bistable regimes of operation [19]. More recently, the protein tyrosine phosphates (PTP) SHP-1 was shown to be critical for defining the ligand discrimination threshold of the ERK response in T lymphocytes [20]. Nonetheless, in view of the multiple phosphatases known to be associated with components of the MAPK pathway, a more integrated view of how such phosphatases regulate input/output relationships is presently lacking[21].

Our earlier study in murine B lymphoma A20 cells had demonstrated that the B cell antigen receptor (BCR)-dependent phosphorylation profiles of all the three constituents of the MAPK module (i.e. Raf. MEK-1/2 and ERK-1/2) were profoundly influenced when cells were depleted of a range of cellular phosphatases by siRNA [22]. Interestingly, depending upon the phosphatase depleted, these effects were either positive or negative suggesting diverse modes of regulation. Therefore, in the present study, we integrated this experimental data with existing literature to generate a mathematical model for ERK phosphorylation. The resulting model revealed an additional level of regulation of the MAPK pathway, which was enforced through the parallel alignment of a cascade of phosphatases. Importantly, the crosstalk between the MAPK and the phosphatases cascade led to the assembly of a novel regulatory motif that enabled independent calibration of signal at each successive node. As a result signal transmission through the MAPK pathway was non-linear, leading to a combinatorial expansion of the landscape of potential output responses. Significantly, in addition to parameters such as signal strength and duration, the bounds of this landscape also incorporated variations in the relative levels (or, activities) of the associated phosphatases. Thus, our studies identify a novel processing principle as a result of which the signal output represents an integrated expression of the properties of the stimulus, and the phenotypic status of the cell.

**RESULTS.**

**Modeling BCR-dependent activation of the MAPK network.**
Our previous experimental studies of the plasticity of the BCR signaling network, and its response to specific perturbations [22-24], provided us with an internally normalized dataset for modeling salient features of the signaling network. We, therefore, combined these results with data from the literature to develop a block diagram for the BCR-activated MAPK signaling network. Here we specifically included the results from our most recent experiments probing the consequences of siRNA-mediated depletion of various cellular phosphatases, on BCR-dependent phosphorylation of the constituents of the MAPK pathway [22]. Data generated from the depletion of five of the tested phosphatases were taken for this study, and these five phosphatases were: PP1, PP2A, MKP1, MKP2, and MKP3. The selection of these phosphatases was guided by the fact that MKPs are known regulators of phoshorylation of the intermediates in the MAPK pathway [19,25,26]. Further Raf, MEK, MKP1 and MKP3 are all regulated through phosphorylation at Ser/Thr residues [25,27], thereby implicating a possible role for PP1 and PP2A.

Modeling of BCR mediated signal transduction involved the basic reaction scheme from the receptor, to the intermediate molecules leading to the MAPK pathway and, eventually, to the regulation of ERK (a detailed description of the model development, parameter estimation, and sensitivity and robustness analyses are provided in Supplementary Data S1) (Fig1A). This process involved several rounds of iterations before the model could completely reproduce the experimental dataset, which involved upstream kinases and phosphatases apart from the core MAP kinase module. Model parameters such as initial concentrations, Km and Kd values, and other rate parameters were either taken from literature or, when not available, were optimized for producing the best experimental fit. The model incorporated generic rate laws such as mass action and Michaelis-Menten reaction schemes to represent the interactions among the species. The system was taken as appropriately modeled when it could fit the experimental phenotypic data (e.g. the previous results of our siRNA experiments), which included both the upstream and the MAPK intermediates. Apart from the general schematic (Fig 1A) of BCR signal transduction leading to ERK, the core MAPK module regulated by cellular phosphatases (Fig 1B) was of our interest for analysis. Care was taken to incorporate these phosphatases in the network in a manner that was consistent with the known literature information and, in cases where adequate information was not available, assumptions were made based on our own experimental data to fit the phosphatases. Overall, the model consisted of 158 variables representing different pools of molecules, 172 biochemical reactions, and 256 total parameters (see Supplementary Data S1).

The resulting model was fairly complex, encompassing the span of regulatory events emanating from the receptor and terminating at ERK (Fig. 1A &B). To test the model, we performed Local and Global Sensitivity analyses using several independent methods to evaluate the parameters for their robustness against fluctuations that are inherent to noisy biological conditions. We found that parameters that were most sensitive were those that are already known to be critical for shaping the BCR-specific signaling response. In other words, our model indeed reflected the true kinetic properties of BCR-dependent signaling. Parameters specific for the MAPK-phosphatase module (see Fig. 1B) were robust in terms of the output behavior, and could tolerate a broad range of perturbations against the ERK activation response. Further, we also verified that the robustness of our model compared well with that of other prominent models described for the MAPK pathway (see Supplementary Data S1). At the experimental level, the model also successfully recapitulated results available in the literature, which in-cluded our own earlier experimental data on BCR-dependent signaling in A20 cells (Supplementary Data S2 and S3).

**Analysis of the model.**
The model described in Figure 1A exhibited several intriguing topological and dynamical features. This is especially evident from Figure 1B, which highlights the co-alignment of a phosphatase cascade alongside the MAPK pathway, leading to several cross-regulatory interactions between the two. Notably, these interactions also included two novel links for MKP3. That is, in addition to ERK, our model described MKP3 to also interact with – and negatively regulate - MEK and MKP1 (Fig. 1B). Further, another related prediction by the model was that MKP3-dependent regulation was tightly controlled through a modulation of its substrate bias. In other words, while phosphorylated MKP3 was described to exhibit a preference for MEK and MKP1, its non-phosphorylated form was predicted to be more selective for ERK (see Supplementary Data S1). As a result, shifts in the distribution of MKP3 between the phosphorylated and non-phosphorylated pools potentially served as a mechanism for diversifying the influence of MKP3 on MAPK signaling.

An *in silico* prediction of the time-dependent profile of ERK phosphorylation was consistent with the experimentally obtained results (Fig. 2A). Further, a similar *in silico* analysis of ERK phosphorylation in response to varying ligand concentrations yielded an incremental dose-response response profile (Fig. 2B). This prediction of a graded ERK output could also be experimentally confirmed by stimulating A20 cells with increasing concentrations of anti-IgG (Supplementary Data S4). Importantly, ERK response to ligand dose was found to be graded regardless of whether it was monitored at the level of the cell population by Western blot analysis (Fig. 2C), or at the level of single cells through intracellular staining for phospho-ERK and detection by flow cytometry (Fig. 2D).

**Experimental verification of the model.**
We have already noted earlier that our completed model for the BCR-dependent activation of ERK retained its ability to describe the activation profiles of the intermediates in a manner that was consistent with the experimental data. Further, predictions on both the kinetics of ERK phosphorylation, and on the graded nature of the ERK response could also be subsequently borne out by experiment. Collectively these results would seem to support the veracity of our model, as well its sensitivity to perturbations in a context dependant manner. Nonetheless, it was also necessary to additionally confirm some of the topological features of the model, and their contribution to the systems properties of the MAPK module.

As shown in Figure 1B, a significant aspect of our model was the description of MKP3 as an important regulator of MAPK signaling through its ability to regulate MEK, ERK, and MKP1. While the interactions between MKP3 and ERK are well known [28], the MKP3-MEK and MKP3-MKP1 interactions represent novel links identified here. We, therefore, sought to verify the existence of these latter links by performing co-immunoprecipitation experiments from lysates of A20 cells, using antibodies specific either for MEK, MKP1, or MKP3. As shown in Figure 3A, immunoprecipitation of either MEK or MKP1 also led to the simultaneous enrichment of MKP3, whereas immunoprecipitates of MKP3 were enriched for both MEK and MKP1. Although immunoprecipitation results do not permit quantification they, nonetheless, support at least the existence of MKP3 in a simultaneous complex with MKP1 and MEK, thereby confirming the interactions depicted in Figure 1B. In subsequent experiments we observed that the siRNA-mediated depletion of MKP3 resulted in a significant (~ 4-fold) increase in the basal levels of MKP1 (Fig. 3B and Supplementary Data S4). These levels were further increased upon stimulation of cells through the BCR (Fig. 3B). The stability of intracellular MKP1 is dictated by its phosphorylation status. While the non-phosphorylated form exhibits a short half-life, phosphorylation enhances its stability [25]. Thus the observed inverse relationship between MKP3 and MKP1 levels in Figure 3B provides additional support for the functional nature of the link between these two phosphatases. In this connection it is pertinent to note that under condition of MKP3 silencing, the level of both phosphorylated ERK and p38 are significantly reduced following receptor stimulation [22]. Thus, the increased stability of MKP1 observed here, following MKP3 silencing, is more likely due to reduced de-phosphorylation of MKP1 by MKP3.

To verify whether the phosphorylation status of MKP3 indeed modulated its substrate-specificity, we also examined anti-IgG-mediated stimulation of A20 cells in the presence of the CK2α inhibitor, DRB [29]. As shown in Figure 1B, MKP3 is primarily phosphorylated by CK2α. Inhibition of this kinase, therefore, leads to a concomitant inhibition of MKP3 phosphorylation [27]. According to our model then, this should then also result in an increased efficiency of MKP3-mediated dephosphorylation of ERK. As expected, stimulation of DRB-treated cells led to inhibition of BCR-dependent phosphorylation of ERK (Fig. 3C and Supplementary Data S4). These results are also consistent with previous findings that CK2α dependent phosphorylation of MKP3 attenuates its inhibitory effect on ERK phosphorylation [27].

Although the results in Figure 3C provided experimental support for our proposal that MKP3 displays differential substrate specificity depending upon its phosphorylation status, we wanted to further establish this using an alternate and more direct approach. For this we again stimulated A20 cells through the BCR, but either in the presence or absence of the CK2α-inhibitor, DRB. Following this, we then employed confocal microscopy to determine the effects of CK2α-inhibition on the co-localization of MKP3 with ERK, MEK, and MKP1. Figures 3D and E show that, in stimu-lated cells, inhibition of CK2α resulted in a substantial increase in the extent of co-localization between MKP3 and ERK. In contrast, co-localization of MKP3 with both MEK and MKP1 was significantly reduced (Figure 3D & E). Essentially similar results were also obtained in experiments where, instead of inhibition, CK2α was also specifically silenced through targeted siRNA (not shown). At one level these results provide direct evidence for the novel links between MKP3 and MEK, and between MKP3 and MKP1 revealed by our model. In addition to this, however, the observed alterations in the relative extents of co-localization of MKP3 with ERK, MEK, and MKP1 – induced by the inhibition of CK2α-dependent phosphorylation of MKP3 - also provide clear experimental confirmation for our proposal that the substrate specificity of MKP3 is influenced by its phosphorylation status.

Collectively then, both the consistency between predicted and experimentally obtained phosphorylation profiles of the various signaling intermediates described above, as well as the experimental confirmation of the new MKP3-dependent regulatory links proposed, provide experimental support for the model depicted in Figure 1A.

**Signal-dependent assembly of the MAPK cascade into a unique regulatory module.**
According to the model in Figure 1B, each intermediate in the MAPK pathway was under the regulatory control of both the co-aligned, and its immediately upstream, phosphatase. Thus, MEK was regulated by both PP2A and MKP3, whereas ERK was regulated by MKP3 and MKP1 (Fig. 1B). The entire model consisted of links that collectively belonged to all of the four possible categories; activation of activator (PKC to Raf, and Raf to MEK), inhibition of activator (PP2A to Raf, and MKP3 to MEK), activation of inhibitor (PKC to MKP3, and ERK to MKP1), and inhibition of inhibitor (PP2A to MKP3 and MKP3 to MKP1, 2) (Fig. 1A and 1B).

A particularly striking aspect here was that co-alignment of the two cascades resulted in the generation of a novel regulatory motif that was composed of a contiguous set of two square units. The first of these consisted of Raf, MEK, PP2A, and MKP3 as the vertices, whereas the second unit involved MEK, ERK, MKP3, and MKP1 (Fig. 1B). Further, the linkage relationships between the corresponding vertices were also replicated (Fig. 1B). Interestingly, a closer inspection of this motif revealed that it was in fact composed of four smaller regulatory units (each triangle in the motif) that represented alternately placed type-2 coherent and incoherent feedforward loops (FFLs, defined as in [30]. A more detailed analysis of these sub-structures, and their individual contributions to the overall properties of the parent motif is currently under way as a separate study. In this report we focused solely on the role of the larger motif, in regulating signal processing by the MAPK cascade.

Assembly of this regulatory motif required two separate input signals originating from the BCR. The first of these was a PKC- and RasGTP-dependent AND gate that mediated Raf activation in a signal-dependent manner [19,31]. The second input signal was defined by the CK2α-dependent phosphorylation of MKP3 (Fig. 1B). It has previously been demonstrated that MKP3 and CK2α form a protein complex wherein CK2α specifically phosphorylates MKP3, with the corresponding increase in ERK-2 phosphorylation [27]. As discussed later, the regulation of MKP3 between its phosphorylated and non-phosphorylated states has significant implications for signal processing by the MAPK pathway. Finally, phosphorylation of MKP1 by ERK was also a prerequisite for the assembly of this motif, signifying the importance of crosstalk between these two cascades in this process.

**Phosphatase-mediated regulation of the ERK response.**
We first examined how the connected network of phosphatases influenced signal output from the MAPK module. This was achieved through *in silico* experiments examining the ligand dose-dependency of ERK phosphorylation, under conditions where the individual phosphatases were depleted one at a time. Figure 4 shows the results of these experiments where panel A depicts the profile obtained in the unperturbed condition. It is evident that, with the exception of PP1, no other phosphatase depletion had any significant effect on the proportional nature of ERK activation (Fig. 4), The observed effect of PP1 depletion is consistent with our earlier experimental findings involving depletion of this enzyme by siRNA. In these experiments, a marked decrease in the magnitude of ERK activation was obtained at the later time points [22]. Thus these results indicate that PP1 may play an important role in buffering the inactive pool of hyper-phosphorylated Raf, which is generated by activated ERK through a positive feedback loop, and Akt [19,32,33].

Depletion of any of the remaining phosphatases resulted only in a modulation of the amplitude of ERK phosphorylation, with MKP1 depletion yielding the most pronounced effect (Fig. 4). Notable, however, were the qualitatively distinct effects seen upon depletion of the individual, MAPK-associated, phosphatases. Suppression of PP2A expression yielded a marginal effect on the shape of the dose response curve (Fig. 4), whereas that of MKP3 resulted in a diminished activation window for ERK due to a simultaneous increase in sensitivity to lower ligand concentrations, and a decrease in the peak level that was achieved (Fig. 4). In contrast, removal of MKP2 induced a comparable increase in both the upper and lower bounds of ERK activity (Fig. 4). Thus, selective depletion of each of the phosphatases exerted its own individual effect on the ligand-dependent dose-response profile of ERK activation.

**Perturbation in phosphatase concentrations modulates the MAPK response.**

The output of a signaling network is primarily determined by the topology of the network, and the concentrations of its nodal constituents. As shown in Figure 4, the topological allocation clearly had a significant bearing in terms of defining the extent of regulation that each phosphatase exerted on the MAPK signaling pathway. We, therefore, next examined how variations in the concentration of these phosphatases would impact on BCR-initiated MAPK signaling. This involved *in silico* experiments in which the concentration of the individual phosphatase was varied from one that was ten-fold lower, to one that was ten-fold higher than its constitutive level in A20 cells. The phosphorylation of both MEK-1/2 and ERK were examined under these conditions.

While such experiments were also carried out for PP1 and MKP2 (see Supplementary Data S5), only the results obtained for the three phosphatases that are incorporated in the 'two-square' structural motif are shown in Figure 5. The figure describes the effect of varying the concentration of these individual phosphatases, on the peak levels of MEK and ERK phosphorylation. Of particular note here was the fact that changing PP2A concentrations exerted dissimilar effects on MEK and ERK. A bimodal effect was observed for MEK and this feature was retained at all strengths of signal (i.e. ligand doses) employed (Fig. 5). Contrary to this, ERK yielded a PP2A-dependent bimodal profile only at low ligand concentrations while the response to this phosphatase became linear as the signal strength was increased (Fig. 5). Thus, PP2A-mediated regulation imparted non-identical effects on these two – consecutively positioned - signaling intermediates of the MAPK pathway. Further, the extent of deviation in behavior between these two intermediates was also dependent upon the strength of the stimulus at the BCR. Increasing PP2A concentrations also led to a progressive reduction in the amplitude of both MEK and ERK responses although the effect on the latter molecule was more pronounced (Fig. 5). This aspect could be experimentally verified by stimulating A20 cells in the presence of the PP2A inhibitor okadaic acid (OA). Although OA is a known inhibitor of both PP2A and PP1, the 100-fold lower $IC_{50}$ value for PP2A allowed us to use a concentration range that was specific for PP2A.

As shown in Supplementary Data S6, while inhibition of PP2A activity led to a linear increase in ERK phosphorylation, that of MEK shows a bimodal response to PP2A (also Figure 7A). The sensitivity of ERK to PP2A activity likely derives from the concomitant decrease in the size of the pool of non-phosphorylated MKP3, as a result of decreasing PP2A levels. As discussed earlier, this would decrease the efficiency of MKP3-dependent dephosphorylation of ERK. In addition to this, however, a diminished pool of phospho-MKP3 would also imply a consequent reduction in the negative regulation of MKP1 and, thereby, an increase in the MKP1-dependent inactivation of ERK.

In similar experiments where MKP3 concentrations were varied, we observed an inverse relationship between the amplitude of MEK phosphorylation and MKP3 concentrations. In contrast to this, ERK activation displayed a bell-shaped profile (Fig. 5). The curvature of the bell was, however, more pronounced at higher ligand concentrations supporting that ERK activation remained responsive to the strength of the stimulus. The bimodal effect of MKP3 on ERK activation again probably reflects the consequence of concentration-dependent shifts in the distribution of MKP3 between its phosphorylated and non-phosphorylated states. At low concentrations, the non-phosphorylated form of MKP3 would predominate due to the dominant action of PP2A. The increased specificity of this form for ERK can then explain the reduced levels of peak ERK phosphorylation. As MKP3 levels increase, however, the proportion of its phosphorylated form is also expected to increase. The consequences of this on ERK activation would then be defined by the balance of the effects of MKP3 on ERK, and that of phospho-MKP3 on MEK and MKP1 respectively. Thus, MKP3 negatively regulates ERK activity by ensuring its dephosphorylation. In contrast, phospho-MKP3 preferentially targets MEK and MKP1 but with opposing consequences. While MEK inactivation also inhibits transmission of signal to ERK, the dephosphorylation and consequent destabilization of MKP1 diminishes the negative regulation of ERK by this enzyme. It is the quantitative shifts in balance between these distinct activities that would then shape the ERK response to variations in MKP3 concentrations.

Consistent with expectations, increasing MKP1 levels led to a steep decline in peak levels of ERK phosphorylation whereas the effect on MEK was only marginal (Fig. 5). Thus the cumulative results in Figure 5 emphasize that signal-dependent formation of the "two-square" motif enables tight regulation of the MAPK signaling by the associated phosphatases. Importantly, both quantitative and qualitative aspects of signal transmission between two consecutive nodes were selectively influenced, thereby, providing a mechanism by which signal transmission at the MEK/ERK interface could be further calibrated.

**The phosphatase axis orients signal processing by the MAPK module.**

Concentrations of phosphatases frequently vary from one cell type to another. Further, even within a given cell type, either differences in environmental conditions (histories of stimulus) and/or developmental stages can lead to significant changes in intracellular phosphatase concentrations [26,34,35]. This is particularly true for MKP's in lymphocytes, where changes either in the activation or differentiation state of the cell can lead to marked alteration in the relative levels of this class of phosphatases [26]. Thus, given the high connectivity between the MAPK and the phosphatase cascade, it was reasonable to suspect that variations in the relative concentration or activity of the MAPK-associated phosphatases would have a significant impact on signal processing. To explore this we examined the consequences of simultaneously varying the concentration of any two of the three MAPK-associated phosphatases, while keeping the third at its constitutive level. The effects on the magnitude of peak phosphorylation of both MEK and ERK, at a fixed ligand dose, were monitored. All the three possible phosphatase combinations were tested, over a range of 10,000 different pairs – in a 100 x 100 matrix – of their respective concentrations. The range of phosphatase concentrations employed here was identical to that described for Figure 5.

Figure 6A depicts the results of such an exercise in which PP2A and MKP3 concentrations were varied, while keeping MKP1 at a fixed level. As shown, MEK activity yields a bell-shaped curve in the direction of increasing PP2A levels, although this profile tends to progressively 'flatten' as MKP3 concentrations also increase. While the inhibitory effect of MKP3 on MEK activation was monotonic it was, nonetheless, buffered at high PP2A concentrations (Fig. 6A). In other words, the relative concentrations of PP2A and MKP3 have a significant role to play in defining the sensitivity of MEK phosphorylation, to ligand-induced stimulation of cells. This was also true in the case of ERK although, here, a bimodal activation profile was obtained in the directions of both increasing PP2A and MKP3 concentrations. This latter observation suggests the existence of compensatory effects between these two phosphatases, whereby multiple combinations of their respective concentrations yield the same level of peak ERK activity.

Figure 6B shows the results of a similar experiment where the concentrations of MKP3 and MKP1 were varied, under conditions where PP2A levels were kept constant. As may be expected, variations in MKP1 concentrations exerted little or no effect on peak MEK activity and the resulting landscape was heavily biased in favor of the concentration-dependent inhibitory effects of MKP3. In contrast, a bimodal activation curve was again observed for ERK along the MKP3 axis. The curvature of this profile, however, was stringently controlled by MKP1, with ERK activation being completely inhibited even at moderately higher concentrations of this phosphatase (Fig. 6B).

Similar to the results in Figure 6A, varying combinations of PP2A and MKP1 also yielded a bimodal MEK profile in the direction of PP2A, but with a significant reduction in the phosphatase concentration window where at least detectable MEK activation was allowed (Fig. 6C). With the exception of attenuating the amplitude of MEK activation, varying MKP1 levels had little effect on the profile. Qualitatively similar effects were also noted for ERK where increasing MKP1 levels served to further exacerbate the inhibition of ERK activation that was observed as a function of increasing PP2A levels. However, here the inhibitory effect of MKP1 was far more pronounced. (Fig. 6C). Thus, although the MAPK module functions as a proportional response system, these results emphasize that the net output is further specified by the relative concentrations of the individual MAPK-associated phosphatases that are present. In this connection, the assembly of these phosphatases into a regulatory axis is particularly relevant. As a result, variations in the concentration gradient along this axis exercise a combinatorial influence on signal processing by the MAPK pathway.

Finally, yet another intriguing feature revealed from a cursory examination of Figure 6 is the fact that the range of allowed concentration regimens of phosphatases was significantly more restricted for ERK than it was for MEK. This was generally true for all the three pairs of phosphatase combinations studied where the activation landscape for MEK was always larger in area than that obtained for ERK (Fig. 6). These observations suggest that the control exerted by the co-aligned phosphatase axis is further accentuated during transmission of signal from MEK to ERK, the effector molecule of the MAPK pathway.

**Experimental support for a regulatory phosphatase axis.** To confirm the above findings we examined - as a representative case - the consequences of experimentally modulating MKP3 levels, in conjunction with variations in the concentration of active PP2A. MKP3 levels in A20 cells were altered either through the use of specific siRNA, which reduces the amount of the protein by >70% [24], or by stably over-expressing it in these cells to achieve a net two-fold increase in its concentration (Materials and Methods). To modulate the concentration of active PP2A, we again employed its inhibitor OA. Thus, cells expressing either constitutive, reduced (through siRNA), or increased (through over-expression) levels of MKP3 were individually treated with varying OA concentrations, and each of these groups were then stimulated with a range of ligand (i.e. anti-IgG) doses. Peak phosphorylation levels of both MEK and ERK were determined in each case, and the results obtained are presented in Supplementary Data S6 and S7.

We first compared these results with the predicted effects - on peak phosphorylation of MEK and ERK - of varying levels of the individual phosphatases shown in Figure 5B. The panel 'a' in Figure 7A shows an expanded region of these predicted profiles obtained at the higher range of ligand doses, and over the estimated range of either MKP3 concentrations or PP2A activities employed in the present set of experiments (see Supplementary Data S8). Panel 'b' of Figure 7A depicts the experimentally obtained results under these conditions. A high degree of correspondence between the simulated and the experimentally derived profiles is clearly evident here. This was also equally true for peak MEK and ERK phosphorylation profiles in cells stimulated with sub-optimal doses of the ligand. In addition to this, the data obtained in Supplementary Data S6 also served to substantiate the results of our *in silico* experiments describing the effects of simultaneously varying MKP3 and PP2A levels, on the magnitude of ERK phosphorylation (see Fig. 6A). A comparison of results from the two approaches is shown in Figure 7B. Here, panel 'a' shows the results of our *in silico* analysis performed by employing either a high, or a suboptimal concentration of ligand for cell stimulation. The corresponding profiles derived from our experimental data (Supplementary Data S7) are shown in panels 'b' of these figures. Again, a good correlation between results from the two approaches is clearly evident here. This was also true when experimental and predicted results for MEK phosphorylation were similarly compared in terms of a heat map (data not shown).

Thus by verifying the accuracy of the resulting predictions, the data in Figure 7 support the fidelity of the model described in Figure 1A. In addition, these findings also firmly establish the role played by the co-associated phosphatase axis, in terms of further calibrating the stimulus-dependent responsiveness of the MAPK pathway.

**DISCUSSION.**

The remarkable versatility of the MAPK module is evident from the fact that ERK activation can display a wide range of activity profiles depending on the interfaces that regulate signal transmission through this module. Thus depending on the cell type, ERK activation can be ultrasensitive, converting graded inputs into a digital output response [10]. Embedding this pathway within a positive feedback loop can lead to a dramatic amplification of signal, and also endow it with both bistability and hysteresis [19]. In contrast, a negative feedback loop combined with intrinsic ultrasensitivity brings about sustained oscillations in ERK phosphorylation [12,14]. The sensitivity of the MAPK pathway to the cellular milieu is further underscored by the fact that, under certain conditions, it can also function as a monostable system yielding a proportional output response at the level of ERK phosphorylation [36-38]. At least some of the determinants of the nature of input/output relationships identified so far are the concentration of MKPs [19], the concentration of the scaffolding proteins involved [15], and the extent of sequestration of ERK in a complex with its kinase, MEK [16]. In addition to defining the threshold between a switch-like versus a proportional response, these parameters also play a critical role in modulating the amplitude and kinetics of the output signal.

In the present case, the proportional nature of the input/output relationship – despite incorporating a distributive mechanism for the dual phosphorylation of ERK - appears to be combinatorially defined through the relative concentrations/activities of the MAPK pathway-associated phosphatases. Thus, *in silico* exercises involving select modulations in phosphatase concentrations yielded an enhanced slope for ERK activation under conditions where either PP1 or MKP3 levels were significantly increased (Supplementary Data S10). Alterations in levels of the MAPK pathway components had no such effect. Consistent with this is our experimental demonstration that siRNA mediated silencing of these individual phosphatases had a significant effect on the amplitude, the stability, the kinetics, and the baseline levels of ERK phosphorylation. A role for filtering spurious noise by MAPK-associated phosphatases has also been recently discussed [17].

Our finding that phosphatase-dependent regulation of the MAPK pathway involves organization of the phosphatases into a co-aligned cascade was especially interesting. Traditionally, negative regulation of intracellular signal transduction is conceived to occur as a discrete set of localized interactions between a phosphatase and its target molecule, and a cascade organization for phosphatases has not been considered so far [26,39]. Importantly, this co-alignment resulted in the formation of a novel regulatory module that was composed of a contiguous set of two square units with the MAPK constituents and the phosphatases PP2A, MKP3, and MKP1 constituting the vertices of these units. We have termed this novel signaling motif, induced upon BCR-dependent stimulation of cells, as the 'Signal Activated Motif' (SAM).

Assembly of the SAM involved two distinct input signals that, in turn, separately activated the functionally distinct components of the module. The first of these involved the combined action of PKC and RasGTP to activate Raf, thereby initiating signal flow through the MAPK cascade. The second signal, on the other hand, involved the CK2α-dependent phosphorylation of MKP3. This event was critical to the regulatory function of the phosphatase axis. Thus, while BCR activation activated signal flow through the MAPK pathway on the one hand, it also simultaneously embedded this pathway within a regulatory module so that the net signal output could be carefully regulated.

Central to the functioning of the SAM was the distribution of MKP3 between its phosphorylated and non-phosphorylated states where the balance between these two forms was defined by the reciprocal actions of the constitutively active PP2A, and the signal- activated CK2α. This segregation of input signals to activate both the positive and negative regulatory components of the SAM motif provided the mechanism by which filtration and appropriate calibration of the signal output could be achieved. Key to this was the differential substrate bias displayed by phosphorylated and non-phosphorylated MKP3, with the relative proportion of these two subsets being governed by the extent to which the input signal (i.e. activated CK2α) could buffer the action of PP2A. The conditional shifts in the equilibrium between these two pools played a critical role in defining the signal output in terms of the ERK activation profile. In other words, this ability to modulate substrate preferences enabled MKP3 to function as an additional response element for signal discrimination whereby both signal amplitude and sensitivity to ligand concentration could be further tuned.

Our subsequent analysis of how output responses were shaped by alterations in phosphatase concentrations also provided interesting new insights into the properties of the SAM. While MKP3 levels were important in defining the amplitude of ERK activation, it was the MKP1 concentrations that strictly defined the window of ERK activation. That output regulation involved the coordinated action of associated phosphatases was also emphasized in our studies examining the effect of simultaneously varying concentra-tion of two phosphatases. For example, the available levels of the remaining two phosphatases further modulated the effects of varying MKP3 concentrations. This was equally true for PP2A and MKP1, thereby reaffirming the signal-dependent assembly of these phosphatases into a functional axis that intimately regulates signal processing.

The net outcome of the coordinated action of phosphatases was that information flow through the MAPK pathway did not occur in a linear fashion. Rather, signal was differentially processed at each intermediate in the pathway. Consequently, in addition to strength of the stimulus, transmission of signal between successive nodes was also rendered sensitive to the relative concentrations of the associated phosphatases. Thus, for example, conditions yielding a potent activation of MEK did not necessarily translate into a similar profile for ERK. Instead, signal transmission from MEK to ERK was additionally regulated by the activity and/or concentration of the two remaining phosphatase components of the SAM. The resultant combinatorial effects then ranged from a marked amplification, to a complete suppression, of ERK activation. Indeed even the limited investigations performed here reveal how the intricate coupling of a kinase with a phosphatase axis can provide for flexible regimes of regulation where the activity and/or concentrations of the participating phosphatases make key contributions to calibration of the signal output. A direct consequence of this was an expanded landscape of potential ERK responses where a given output characterized not only the strength of activating stimulus, but also the phenotypic state of the cell – at least when characterized in terms of the MAPK-associated phosphatase milieu.

Differences in the tissue-specific origin of cells are also often reflected at the level of differences in concentration of the individual MKPs. Similarly, variable MKP levels frequently characterize the individual stages of lineage commitment in lymphocytes [40]. Further, even within a given cell type, activation of ERK leads to the upregulation of both MKP1 and MKP3 [26], with the consequent desensitization of ERK to any further stimulation of the cell. Thus, the composition of at least MKPs within a cell may well provide a phenotypic description of its activation and/or differentiation status, which then is also incorporated when defining the ERK output response.

Thus, in summary, our present delineation of an emergent motif that endows the MAPK pathway with flexible regimes for modulating signal output provides additional insights into mechanisms that facilitate information processing by the signaling machinery. At one level, these findings reveal a novel organizational principle wherein a kinase cascade is intricately coupled to a regulatory cascade of phosphatases. The resulting structural motif that links each kinase intermediate to both the iso-stage, and the upstream phosphatase ensures that signal transmission is subjected to tight scrutiny at every step of the pathway. This allows the 'strength' of the links between the phosphates axis and the kinase cascade to also function as determinants of signal processing. As a result, even a simple proportional response system becomes endowed with the ability to generate an output that, in addition to signal duration and strength, is also sensitized to variations in phosphatase concentrations and activities. As noted earlier, such variation could potentially encapsulate differences in the phenotypic status of cells either at the level of cell type, differentiation state, or, the history of prior stimulations. It, however, remains to be seen whether such a regulatory module is also associated with the other pathways of the signaling network.

**MATERIALS AND METHODS.**

**Model building and numerical simulations**:

A complete description of model building and details of the reaction scheme is provided in the Supplementary data S1 (Description of model). All reactions were converted into molecule-molecule interactions defining either binding interactions, or catalytic reactions. We have used Simbiology2.2 (Mathworks) on MATLAB7.5 as a platform for implementing the model and for carrying out numerical simulations. All calculations except for that of restimulation experiments were done using ode15s, which is a solver for stiff differential equations and DAE. It uses the variable order method based on numerical differentiation formula for stiff differential equations. Restimulation experiments were done using Sundials solver CVODE provided with Simbiology. Parameters were either taken from literature or estimated. Estimation of unknown parameters was done either by iteratively fitting to the experimental constraints, or by using local optimization with *lsqnonlin, lsqcurvefit* (wherever experimental data was available for immediate readout). Sensitivity analysis of the complete model was done by using SBML_SAT [41], a freely available SBML-based MATLAB toolbox available at (http://sysbio.molgen.mpg.de/SBML-SAT/).

**Stimulation of cells and detection of phosphoproteins**

A20 cells ($1 \times 10^7$/ml) were stimulated with the $F(ab)_2$ fragment of goat anti-mouse IgG at a final concentration of 25 μg/ml in RPMI for a period of up to 30 min [23]. At appropriate times, aliquots of cells were collected, centrifuged, and the cell pellets stored in liquid nitrogen. When required, cells were lysed, the detergent-soluble proteins resolved by SDS–PAGE, and specific proteins and phosphoproteins detected (and quantified) by Western blot using the procedure and antibodies as previously described [24].

**Co-Immunoprecipitation:**
Lysates were prepared from $2 \times 10^7$ cells/group/time point in a buffer containing a cocktail of protease and phosphatase inhibitors as previously described [23]. Lysates were pre-cleared with protein A-agarose and then incubated with appropriate antibody (0.6 – 1 mg, at 4°C for 2h), after which the immune-complexes were precipitated with protein A-sepharose. The immunoprecipitates were identified by immunoblotting with specific antibodies (Santa Cruz Biotech).

**Over-expression of MKP3 in A20 cells.**
MKP3 was over-expressed as a Strep-tag II fusion protein in A20 cells[42]. For this the cDNA was cloned into the pEXPR-IBA103 vector and then transfected into cells by electroporation. Subsequent selection for neomycin resistance over a 4-week period yielded stably-transfected cells. A Western blot analysis revealed that these cells expressed about 2-fold higher levels of MKP3 when compared to untransfected A20 cells.

**Confocal Microscopy: staining and image analysis.**
Immuno-fluorescence staining was performed following the standardized protocol reported previously (Kumar et al, 2008). Stained cells were observed with a Nikon TE 2000E microscope equipped with 60X/1.4NA PlanApochromat DIC objective lens. Secondary antibodies tagged either with Alexa 488 or Alexa 568 were excited at 488 nm and 543 nm with an argon ion- and a Helium-Neon laser respectively. The emissions were recorded through emission filters set at 515/30; 605/75. Serial confocal sections (0.5 μm thick) within a z-stack spanning a total thickness of 10-12 μm were taken in individual channels green and red using the motor drive focusing system. The transmission and detector gains were set to achieve the best signal to noise ratios and the laser powers were tuned to limit bleaching fluorescence. The refractive index of the immersion oil used was 1.515 (Nikon). All settings were rigorously maintained for all the experiments. All images were quantified using Image-Pro Plus version 6.0, a commercially available software package from Media Cybernetics.

**ACKNOWLEDGEMENTS.**

This work was supported by a grant to KVSR from Department of Biotechnology, Govt. of India. VKC is the recipient of a Senior Research Fellowship from the Council for Scientific and Industrial Research (CSIR), and MM the recipient of a Senior Research Fellowship from the Department of Biotechnology, Government of India. We thank Ubaidullah for help with the flow cytometry experiments.

**COMPETING INTERESTS STATEMENT**

The authors declare that they have no competing financial interests.

**FIGURE LEGENDS:**

**Figure 1: Signaling through the Lymphocyte receptor induces a novel regulatory motif for MAP kinase activation.**
A schematic overview of the reactions for BCR dependent activation of MAP kinase ERK-1/2 is shown in Panel A. The highlighted area represents the novel signaling motif identified and this is expanded in panel B.

**Figure 2: The ERK phosphorylation response is proportional to the stimulus strength.**
Panel A shows the concordance obtained between experiment (red diamonds, values as mean ±S.D., n=3) and simulation (black line) examining the time course of ERK phosphorylation obtained upon stimulation of A20 cells with a saturating (25 µg/ml) concentration of anti-IgG. Panel B shows the results of an *in silico* analysis estimating the magnitude of ERK phosphorylation obtained after stimulation of cells with varying anti-IgG concentrations. Panel C gives the corresponding results of an experiment where A20 cells were stimulated for 10 min, with the indicated doses of anti-IgG. ERK phosphorylation was then determined in lysates by Western blot analysis (see Supplementary Data S4). Values are the mean (±S.D.) of three independent experiments. A semi-log plot of these results yielded a Hill coefficient of 0.6 and 0.56 for simulated and the experimental results respectively. Stimulated cells were also subjected to staining for intracellular phospho-ERK using specific antibodies followed by FITC-labeled secondary antibodies. Stained cells were then analyzed by flow cytometry and the results are shown in panel D. Depicted here are the profiles obtained for cells stimulated with either 0.1 (black line), 0.5 (green line), 5 (pink line), or 25 µg/ml (blue line) of anti-IgG. The profile for unstimulated cells overlapped with that for cells stimulated with 0.1 µg/ml of ligand. For the negative control, cells were stained with rabbit IgG instead of the anti-phospho-ERK antibody.

**Figure 3: Verification of novel interactions incorporated in the model**
In panel A, lysates from stimulated or unstimulated A20 cells were immunoprecipitated with antibodies specific either for MKP3, MEK, or MKP1 (Materials and Methods) (left column, IP). Immunoprecipitates were then subjected to a Western blot analysis using the antibodies indicated in the right column (WB). As a negative control (ctrl), parallel immunoprecipitates with rabbit IgG were also probed with the corresponding antibodies.

Panel B shows the results of Western blot analyses for MKP1 in A20 cells treated either with MKP3-specific siRNA (red line), or with non-silencing (i.e. GFP-specific) siRNA (blue line). Stimulation times with anti-IgG (25 µg/ml) are indicated, and values are the mean ± S.D. of three separate experiments. We have previously demonstrated that siRNA-mediated silencing of MKP3 results in a >70% reduction of this phosphatase protein in A20 cells [22].

In Figure 3C the top panel shows the effect of DRB (30µM) on the anti-IgG dependent phosphorylation of ERK. The lower panel shows the corresponding profiles obtained from an *in-silico* analysis. The blue color identifies the profiles obtained in the absence of any inhibitor, whereas the pink color denotes the profiles obtained in the presence of DRB. For the experiment in panel D, A20 cells were stimulated for 10 minutes with anti-IgG in the presence or absence of casein kinase inhibitor DRB (20µM). Cells were then fixed and stained for ERK1/2 and MKP3 (top panel), MEK1/2 and MKP3 (middle panel), or MKP1 and MKP3 (bottom panel) and observed under laser scanning confocal microscope (Materials and Methods). Merged images for co-localization between green (ERK1/2, MEK1/2, or MKP1) and red (MKP3) are shown in the figure. The quantitative differences between DRB treated and untreated cells for co-localization with MKP3-MEK1/2, MKP3-ERK1/2, MKP3-MKP1 upon anti-IgG treatment are shown in panel E. Values are the average (± S.E.) of over at least 40 cells (*** $p < 0.0005$; ** $p < 0.005$).

**Figure 4: Phosphatase-mediated regulation of the MAPK signaling response.**
The influence of individual phosphatases in sensitizing the ERK output was monitored *in-silico* by analyzing ligand dose versus peak phospho-ERK levels within the first 30 min of activation. This analysis was performed either in normal cells (Normal), or in cells where the indicated phosphatase was depleted from the system. In each panel, the fold change in ligand concentration required to increase ERK phosphorylation from 10% to 90% of its maximal value is also given (F). These values confirm that the ERK response remains proportional under all of these conditions.

**Figure 5: Modulation in phosphatase concentrations further shape ligand-sensitivity of the MAPK pathway.**
Peak phosphorylation of ERK (ppERK) and MEK (ppMEK) was monitored in response to variations in both ligand and individual phosphatase concentrations. Three separate sets of simulation, within a defined concentration window, were used to reduce the computing time. For each phosphatase, 100 different concentrations of phosphatase were used, varying from 10 times lower to 10 times higher than its constituent concentration level in A20 cells. The combined data is plotted here. Figure 5A profiles peak phospho-MEK levels as a combined function of both varying ligand doses, and varying concentrations of each of the three phosphatases. Here, the top panel shows the 3-dimensional plot, while the lower panel depicts the same results in the form of a heat map. Similarly, panel B shows peak phospho-ERK levels as a combined function of both ligand dose and varying concentrations of the three phosphatases.

**Figure 6: Inter-phosphatase crosstalk regulates signal transmission through the MAPK pathway.**
The effects of simultaneous variations in the concentration SAM-associated phosphatases on peak phospho-MEK (ppMEK) and peak phospho-ERK (ppERK) levels within a 30-minute activation window are shown here. A total of 100 different concentrations were employed for each phosphatase. These varied from 10 times lower, to 10 times higher than the constituent concentration in A20 cells. The three combinations shown here are MKP3-PP2A (A), MKP3-MKP1 (B) and PP2A-MKP1 (C). In each case the top panel shows peak phospho-MEK levels, while the lower panel shows peak phospho-ERK responses. The left and the right panels depict the 3-dimensional landscape, and the corresponding heat-map representation respectively. The corresponding profiles for the remaining components of the module that are also regulated by phosphorylation (i.e. Raf, MKP1, and MKP3) are shown in Supplementary Data S9.

**Figure 7: Experimental confirmation of the systems properties of the MAPK-associated regulatory module.**
Panel A depicts MEK and ERK phosphorylation levels obtained as a function of changes either in MKP3 or PP2A concentrations as described in the text. For the profiles obtained *in silico* (Predicted), the concentration ranges utilized is described in Supplementary Data S8, whereas the extent of variation in phosphatase concentration/activity obtained experimentally (Experimental) is described in the text. The anti-IgG concentrations employed for the experiment in high ligand dose (upper panel) were 10 (blue line) and 25 (red line) μg/ml; whereas for low ligand dose (lower panel) it was 0.05 (blue line) and 0.5 (red line) μg/ml.

Panels B describes the effects of combined variations in levels/activities of both MKP3 and PP2A, on the magnitude of ERK phosphorylation. Here again a comparison between the *in silico* (Predicted), and the experimentally (Experimental) obtained results is shown. Although a similarity in profiles between the two groups is clearly evident, the absence of a higher degree of concordance was primarily due to the limited number of data points in the experimental group. As described in the text, MKP3 levels were varied in the Experimental sets either through specific depletion by siRNA (KD), or by over-expression (OE).

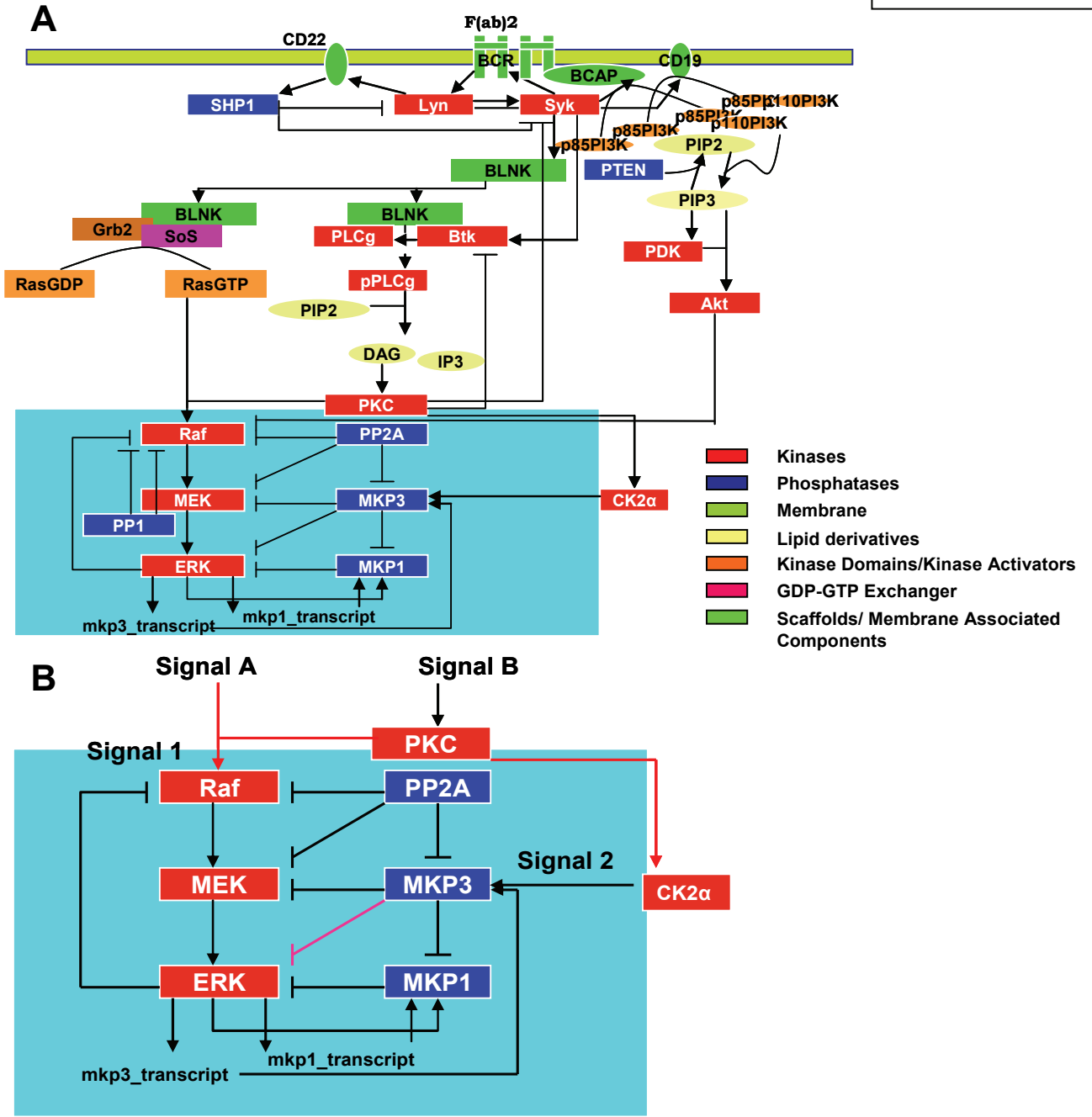

Figure - 1

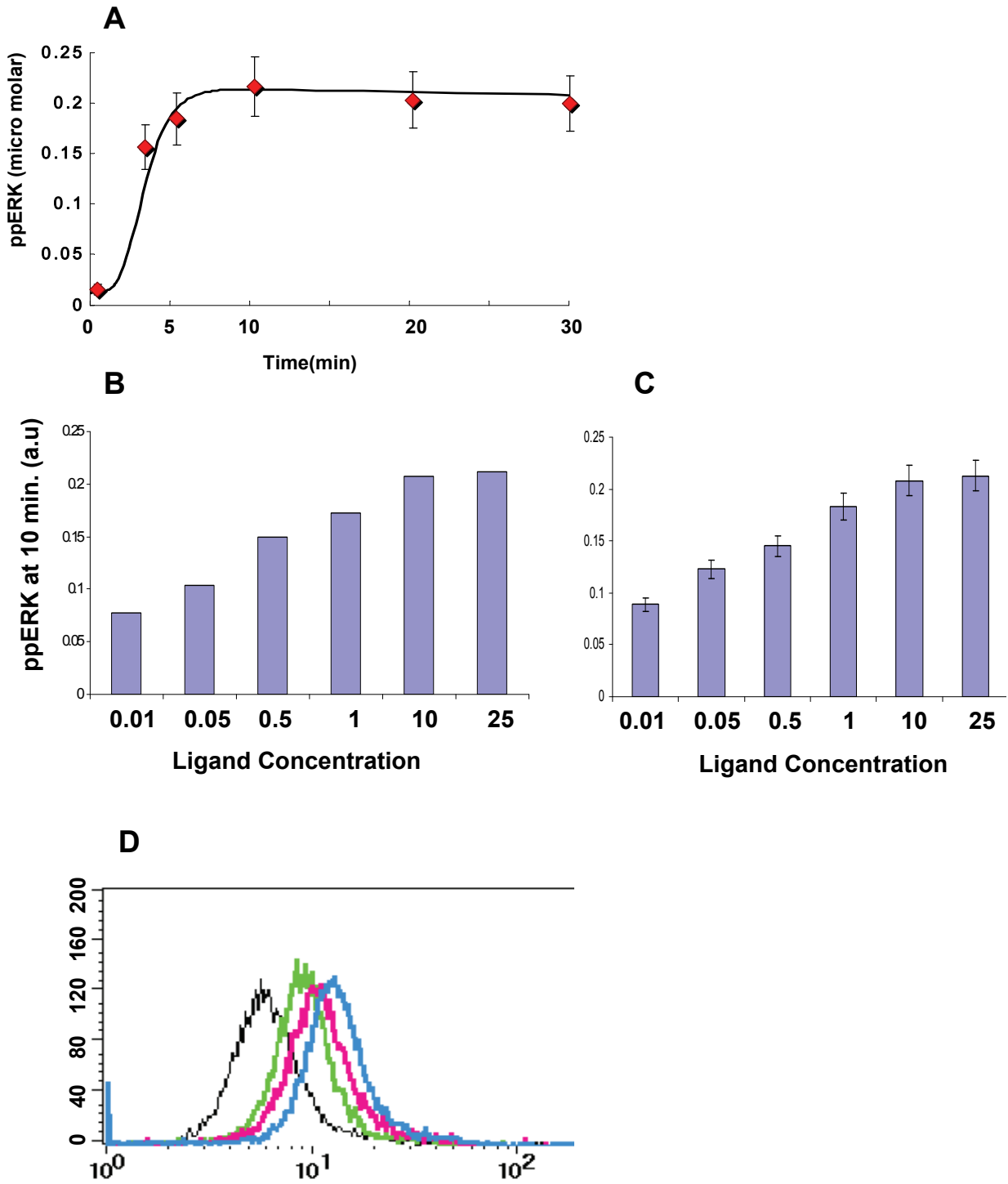

Figure - 2

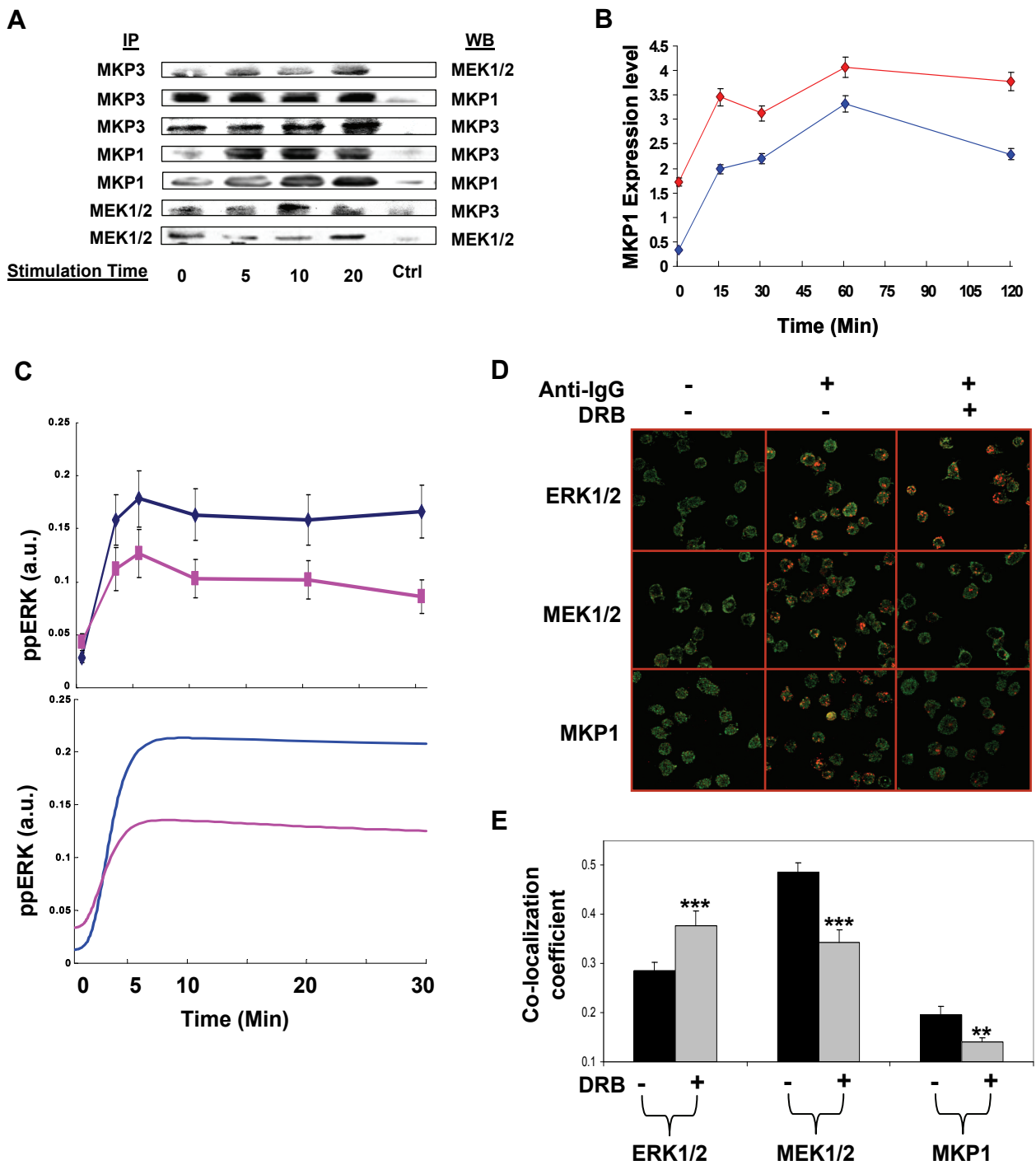

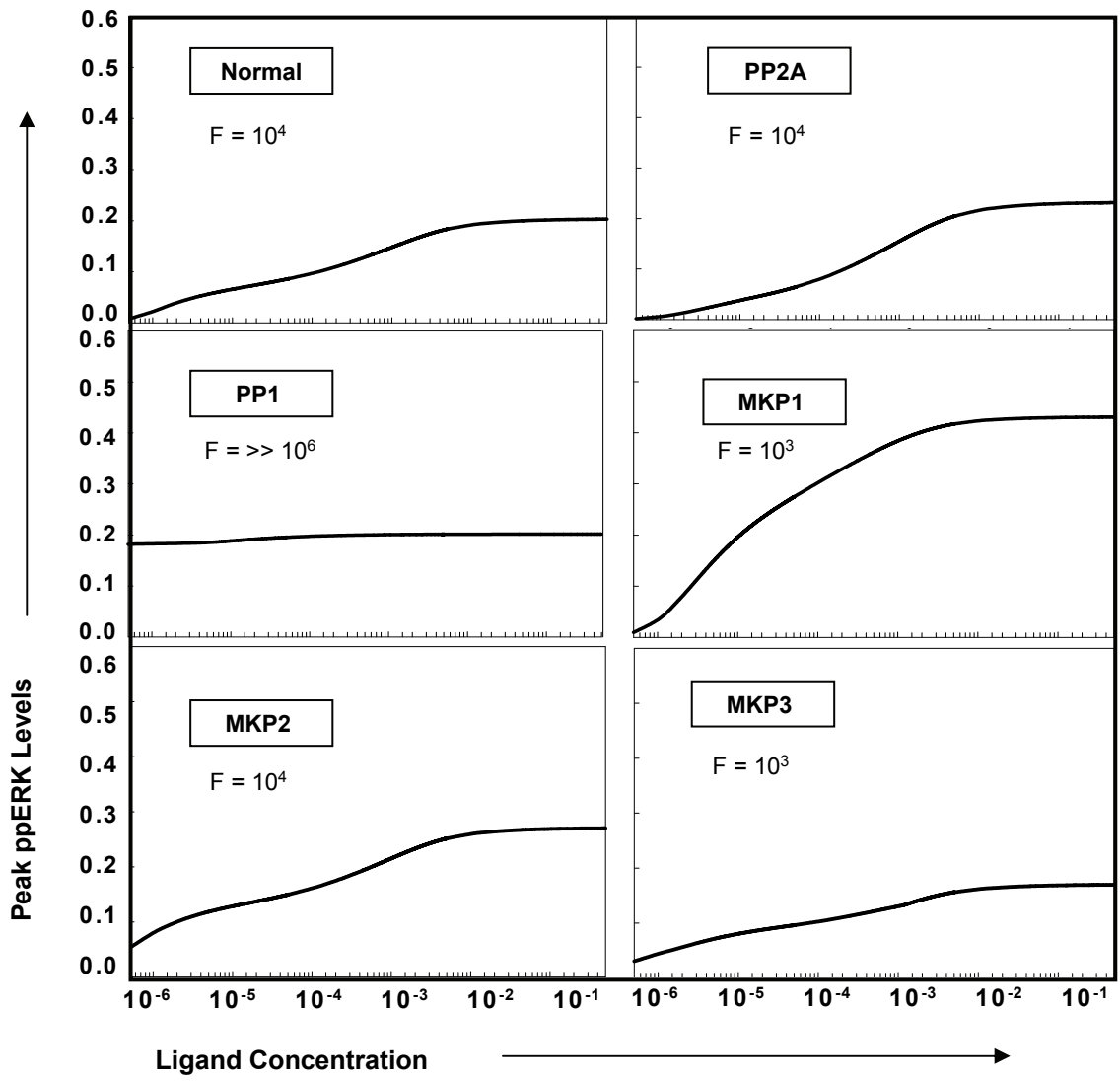
Figure - 4

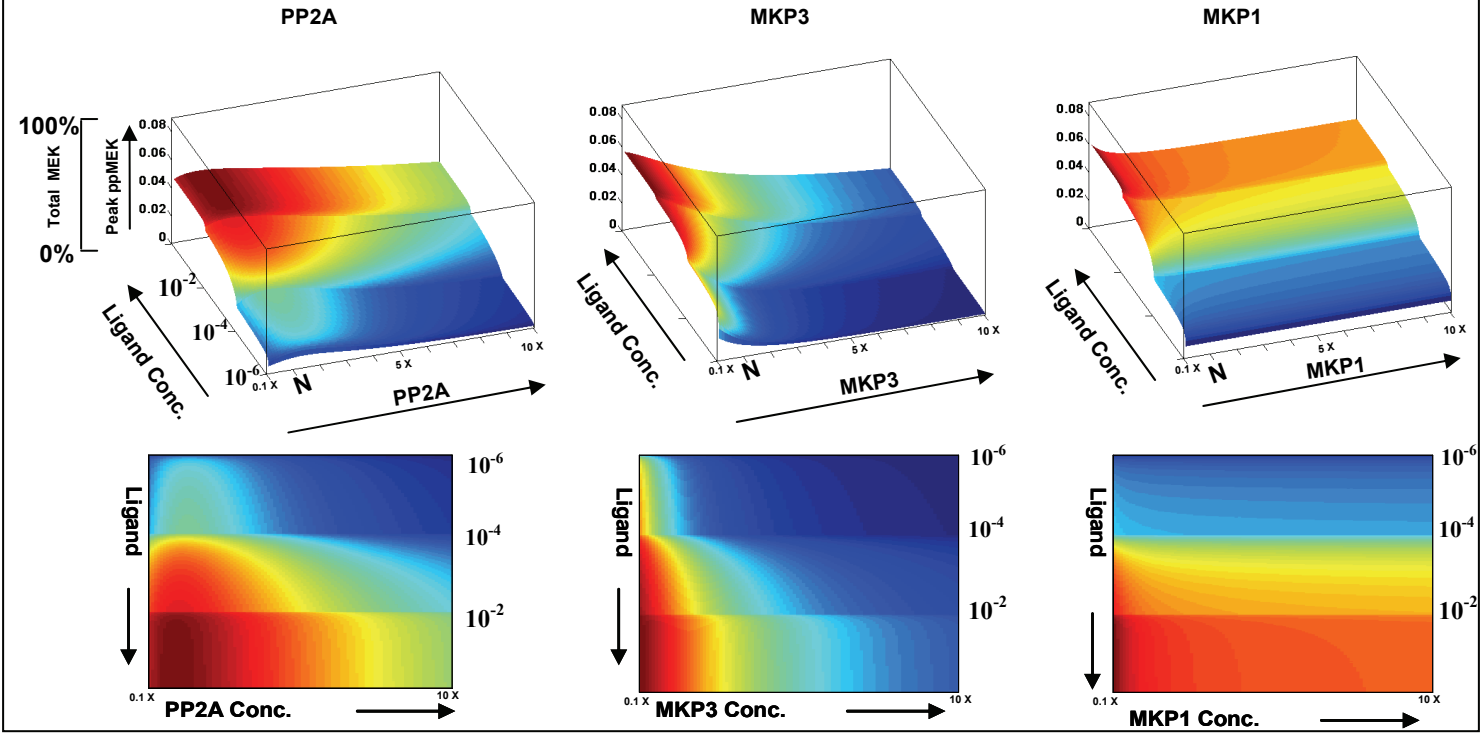

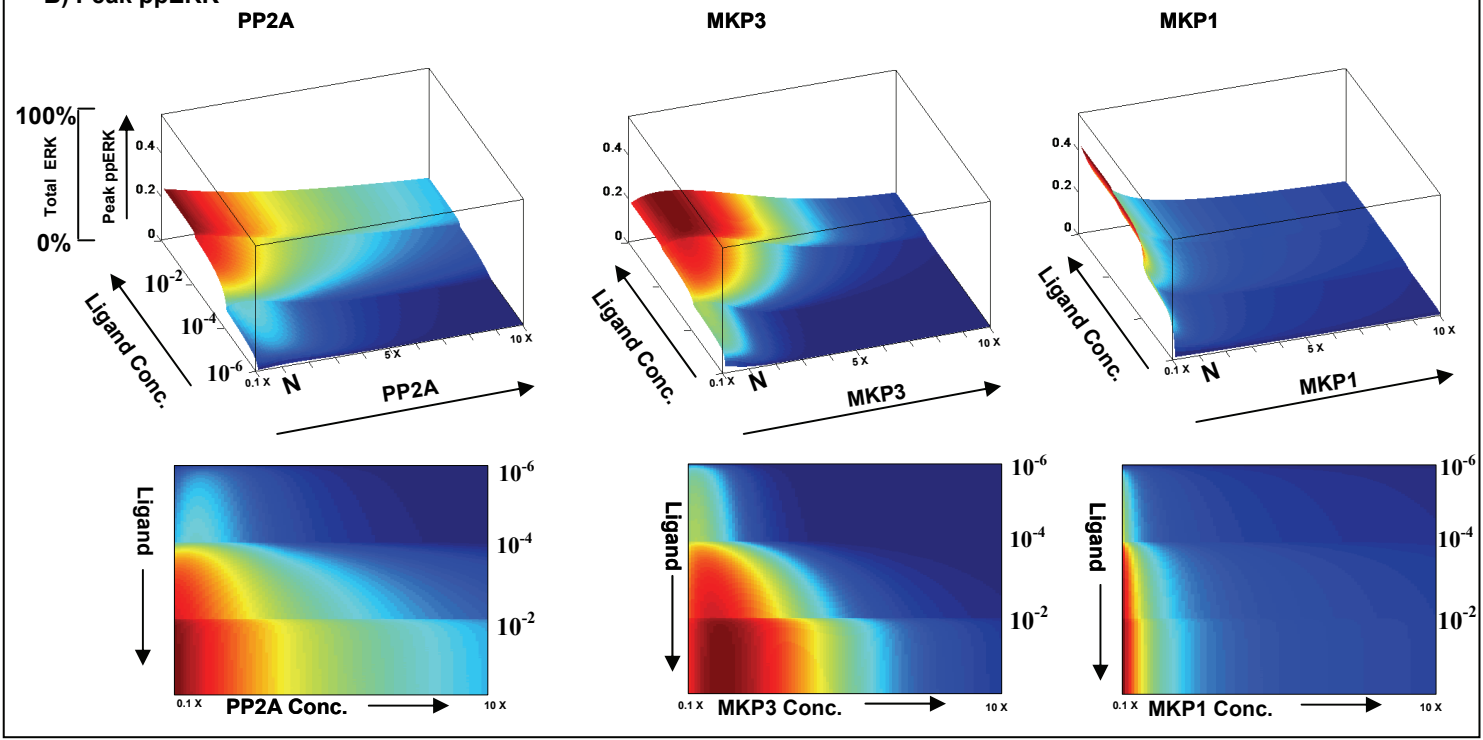

Figure - 5

A) Peak ppMEK

B) Peak ppERK

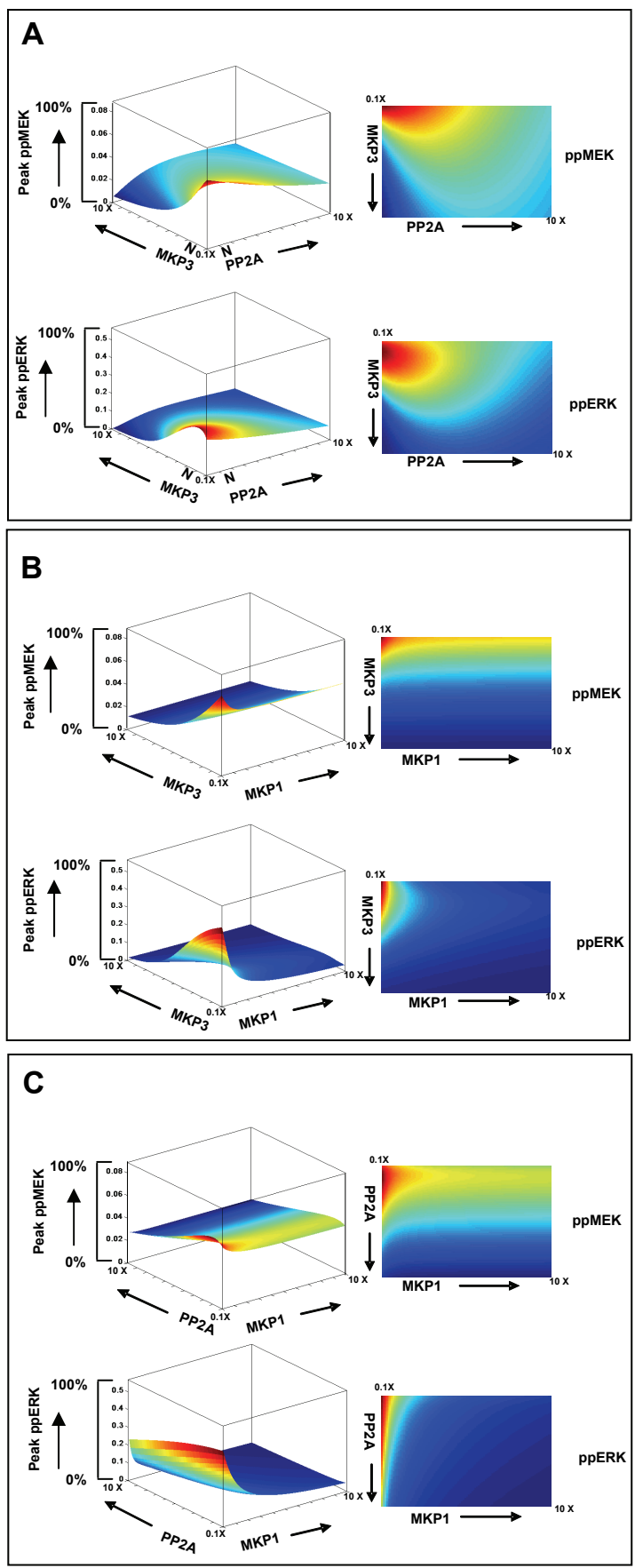

Figure - 6

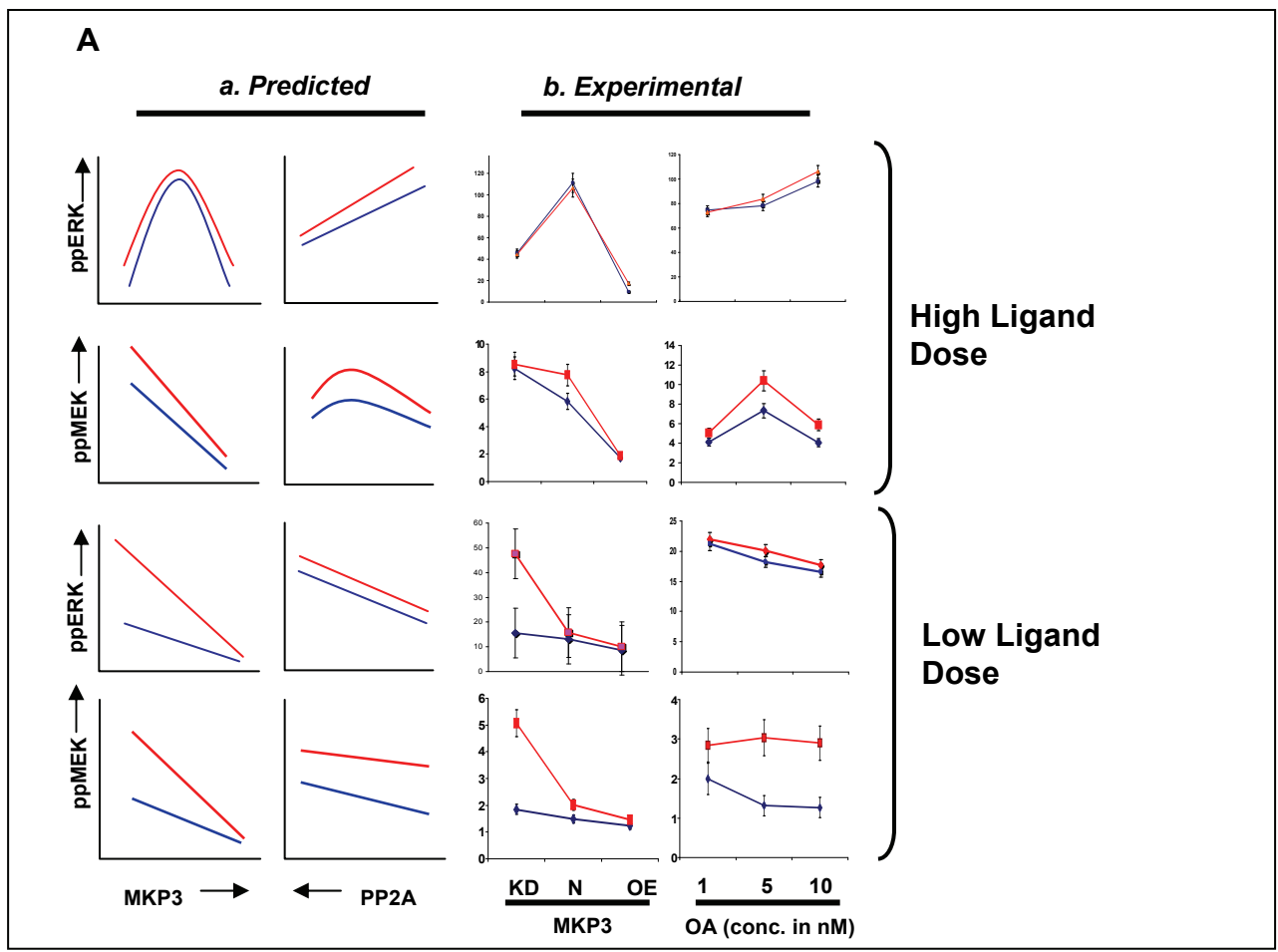
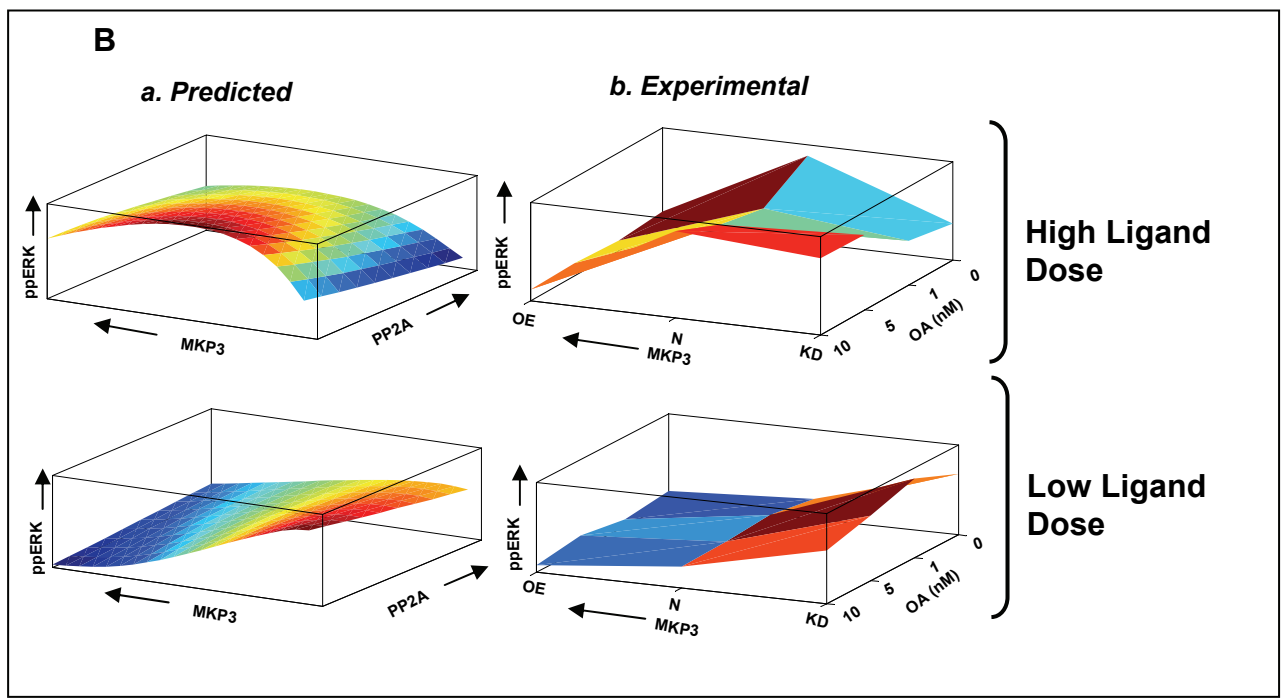

Figure - 7

# Supplementary Information

**Complete description of the Model**

The BCR is composed of heavy and light chains and the associated immuno-tyrosine activation motifs (ITAMs). ITAMs get phosphorylated in response to BCR engagement by antigen, which then leads to cross linking of the BCR molecules. The BCR dependent phosphorylation of ITAMs is sustained through Syk dependent phosphorylation on three additional sites on the ITAMs. For simplicity, however, we have defined all stages of this multiple phosphorylated states as the doubly phosphorylated BCR (which represents- ITAM + Immunoglobulin part of the BCR). Both singly and doubly phosphorylated BCR activates the src kinase Lyn. Syk, which is immediately downstream to Lyn, is then phosphorylated in a Lyn-dependent manner. This establishes a positive feedback loop at the BCR[1]. Antigen-bound BCR also undergoes internalization, although internalized receptors do not contribute towards further signaling[2]. Lyn also phosphorylates CD22, a membrane anchored intermediate of B cell signaling. This, in turn, recruits the protein tyrosine phosphatase SHP1 to the membrane. Since SHP1 is a negative regulator of a majority of protein tyrosine kinases including Lyn and Syk, a negative feedback loop is established. Thus a negative and a positive feedback loop are both established in the immediate vicinity of the BCR. Both Lyn and Syk also activate Brutons' tyrosine kinase (Btk). Btk in turn phosphorylates PLCg when the latter is bound to the Syk-phopshorylated BLNK, a major scaffold protein that connects BCR initiated signaling to downstream events[3,4]. Lyn via CD19, and Syk via BCAP (B Cell Adopter Protein), activate PI3-Kinase signaling with the consequent generation of PIP3. This leads to recruitment of cytoplasmic Akt and PDK to the membrane, where PDK then phosphorylates Akt[4].

Syk mediated activation of BLNK also leads to the formation of a BLNK-Grb2-SoS complex, which functions as the RasGDP-GTP exchanger in B cells. The GAP and GEF are modeled according to the scheme as described by Bhalla et. al.[5,6]. We've summarized the activation of all PKC isoforms by DAG mediated PKC recruitment to the membrane. RasGTP binds to Raf and resulting complex of Raf-RasGTP is then phosphorylated by PKC. Phosphorylated Raf activates the classical Raf-MEK-ERK pathway. Although lymphocyte signaling also involves RasGRPs in activation of RasGTPs via SOS[7,8], the effect of RasGRPs are better characterized in terms of maintaining the basal levels of RasGTP activity[8]. The stimulus induced effect of RasGRPs on RasGTP kinetics is still unclear. Therefore, we kept our model at the level of RasGTP simple and invoked only the classical activation pathway of RasGTP via SoS as the standard GEF. The predicted ligand dose response yielded a good match with the experimental results and more importantly the simplification of the activation pathway of RasGTP did not affect the performance of our model. ERK negatively regulates Raf by hyper phosphorylating it[6,9]. In addition Raf is also negatively regulated through phosphorylation at ser259 by activated Akt [10]. This process has been modeled accordingly. Dephosphorylation of activated Raf is mediated through PP2A whereas the inactivated form represented by hyperphsophorylation or Ser259 phosphorylation, is dephosphorylated by both PP1 and PP2A.

Several reports in the literature have indicated that phosphorylation of ERK1/2 is also regulated by phosphatases[6]. In more recent experiments we have demonstrated that the BCR dependent phosphorylation profiles of all the three constituents of the MAPK module (i.e. Raf, MEK1/2 and ERK1/2) were profoundly influenced when cells were depleted of a range of cellular phosphatases by siRNA. Interestingly, depending upon the phosphatases that were depleted, both negative and positive effects were noted suggesting the existence of diverse modes of regulation [11]. Therefore to explore this further, we also incorporated these earlier results for five phosphatases. These were PP1, PP2A, MKP1, MKP2 and MKP3. We started by linking core MAPK module recursively with the phosphatases in a manner that satisfied the experimental data. The phosphorylation profile of ERK was used as eventual output, although the simulated profiles of Raf and MEK1/2 were also compared with the experimental data for the purpose of model validation.

Whereas PP1(or, Calcineurin) was treated as an unregulated pool, the dual specificity phosphatases MKP1 and MKP3 were modeled as previously described[6], but with the modification that phosphorylation of MKP1 was modeled as a single step process. An interesting aspect of our earlier experimental results was the finding that depletion of MKP3 by siRNA resulted in contrasting effects on the BCR-stimulated phosphorylation profiles of MEK and ERK. Whereas phosphorylation of MEK-1/2 was markedly enhanced, the sensitivity of ERK-1/2 to BCR stimulation was attenuated[11]. This observation could be rationalized by invoking the related finding of [12], that the PKC substrate Casein kinase2α both interacts with and phosphorylate MKP3. Thus, although MKP3 is known to negatively regulate ERK1/2 activation, the simultaneous over-expression of CK2α – however reversed this effect. This attenuating effect of CK2α on MKP3 involves phosphorylation of the latter[12]. In our present system we found that MKP3 expression was up regulated upon BCR stimulation (S3c.B) and our model depicts that the interaction of CK2α with this phosphatase modulates the MKP3-mediated co-ordination of the negative feedback loop influencing ERK-1/2 phosphorylation. Importantly our earlier results also implicated MKP1, which is positively regulated by ERK-1/2 as a direct target of MKP3. While modeling the MAPK pathway, however, the possibility that the MKP3 acted simultaneously on both ERK-1/2 and MKP1 seemed unlikely since such a proposition failed to explain the previously described experimental results. We therefore, considered the possibility that the ERK-1/2 and MKP1 may represent alternate substrate for MKP3, with substrate specificity being determined by the phosphorylation state of MKP3. Whereas the non phosphorylated MKP3 functioned as a negative regulator of ERK-1/2 activation, its phosphorylation by CK2α alters its substrate preference in favor of MKP1, with the consequent attenuation of its activity. This postulates yielded a satisfactory solution after a few optimization and simulation exercises.

Our earlier findings that the depletion of MKP3 from cells by si-RNA resulted in a marked enhancement in the BCR-dependent phosphorylation of MEK-1/2 [11] also prompted us to link phosphorylated MKP3 as a negative regulator of MEK-1/2. Consequently, while phospho-MKP3 would not be expected to directly inhibit ERK-1/2 activation this would nonetheless, be compensated by the inhibitory effect of MKP3 on the upstream kinase MEK-1/2. Further, PP2A has been described as a regulator of the phosphorylation of MEK-1/2 [5,6,13]. In addition, a marked increase in ERK-1/2 phosphorylation was also observed under these conditions. We accounted for these unexpected findings by considering the possibility that phosphorylation of MKP3 could be reversed by the direct action of PP2A. Thus in addition to de-phosphorylating Raf and MEK-1/2, PP2A is also likely to function in regulating the balance between the phosphorylated and non-phosphorylated pools of MKP3. The resulting model for the MAPK module yielded a prediction for the time dependent phosphorylation of ERK-1/2 that was in excellent agreement with the experimentally obtained results.

The above results thus identify a cascade of phosphatases (PP2A-MKP3-MKP1) that is aligned in parallel with the MAPK cascade (Raf-MEK-ERK). ERK activation also leads to transcriptional up-regulation of MKP1 and MKP3. We have achieved the same through a reaction scheme similar to as proposed by [6], although the rates are scaled according to the present concentration of these two phosphatases. The concentrations of c-Raf, MEK, ERK, PP2A, MKP1, RasGDP-GTP total, Akt were determined by Western blot titrations that compared against corresponding levels present in NIH3T3 and CHO cells (Data not shown here). Concentrations of the remaining molecules were estimated.

**Assumptions and criteria for implementation of the model :**

Few important Considerations have been taken while implementing model in mathematical form which are as follows:

1. Since a detailed description of the early signaling events at the BCR signalosome is currently unavailable, the steps were minimized and unnecessary complexity was avoided. For this, the parameters were constrained to the extent possible for describing the kinetics of these events without affecting the known outputs, and keeping the system simple. For example, phosphorylation of BLNK at multiple sites was represented by a single step. Further, the four different (1+ 3) events of phosphorylation at the ITAMs (BCR in the model) by BCR activation and Syk mediated feedback were simplified in two steps (1 by BCR engagement + 1 by Syk) keeping the positive feedback effect of Lyn-Syk-ITAM feedback loop intact.
2. Activated to unactivated ratios were maintained for different molecules in accordance with their cytoplasmic locations and mode of activation. Compartmentalization was not considered in our analysis.
3. In the case of regulation by multiple phosphatases, the experimental results obtained by Western blot analyses cannot be directly correlated to exact relative levels of activated molecules since quantitation by this technique is not linear over a large range. Thus a true regression scheme cannot be employed for phosphatase-mediated perturbations involving large variations in phosphorylation levels. Further, shifts in baseline phosphorylation levels also cause problems, which are not usually encountered in the traditional model building exercises where maximum phosphorylation levels of substrates following on receptor activation are considered as 100%, and regression is then applied for a comparatively smaller range of modulations in pholevels. However post implementation techniques were employed for assessing regression stability of the parameters. These were Multi Parametric Sensitivity Analysis (MPSA) and Partial Rank Correlation Coefficients (PRCC), as described later in sensitivity analysis and robustness scores.
4. Parameters describing the kinetics of molecules on the network periphery were grouped together to define the representative kinetic behavior relevant to the present MAPK module.

**Robustness of the Model:**

Our present model was unique in that it also modeled active regulation by phosphatases. It included active regulation of phosphatase by the kinase components of the network, which in turn was negatively regulated by these active phosphatases. Thus it became important to test the model for its robustness to the parameter variations, at least to the extent that it compared well with existing models. Robustness is a fundamental property of biological systems, which allows the systems to maintain its behavior against random perturbations [14,15]. We preformed the Robustness analysis by SBML_SAT [16]. Robustness analysis was performed as described [17,18], and Robustness scores measured as Total Parameter Variations (TPV) by randomly generated parameters by Latin Hypercube Sampling method.

The R scores (Robustness scores) of a biological components of those models, calculated under identical conditions of perturbations, are given in Table S1.2. As is evident here, the robustness score of our model compares well with that of the others, particularly with those that take phosphatases into consideration. Our model shows a particularly good agreement with that of *Bhalla et.al.* (2002); model assumes a negative value and the closer this value is to zero, the more robust the model. Here we compared some of the models for the MAPK pathways in the literature with our present model. The corresponding R scores of MAPK the only other model that incorporates active regulation of the phosphatase MKP1.

**Table S1.2: Comparison of the robust scores of various Models of MAPK :**

| Model Name | Receptor/ Activator | RasGTP | pMAPKKK | ppMAPKK | pMAPK | Phosphatase Regulation |
|---|---|---|---|---|---|---|
| Huang Ferrell | -0.00543320 | ------- | -0.662038 | -0.961861 | -1.23605 | No |
| Kholodenko 2000 | ------- | ------- | -0.537109 | -1.00071 | -1.25602 | No |
| Levchenko With scaffold | ------- | ------- | -0.344902 | -0.67925 | -1.11045 | No |
| Levcheno No Scaffold | ------- | ------- | -0.561329 | -1.3463 | -2.80383 | No |
| Hatakeyama 2003 | -0.58779 | -1.53691 | -1.59975 | -1.78692 | -3.52799 | Phosphatases Buffered |
| Sasagawa | -0.472733 | -0.79791 | -0.875844 (cRaf) -0.980000 (BRaf) | -1.2053 | -1.99484 | Phosphatases Buffered |
| Bhalla 2002 | -1.44197e-034 | -0.158336 | -0.88649 | -3.01403 | -5.13613 | MKP1 Regulation by MAPK |
| **Present Study** | -0.386837 | -0.980019 | -1.34911 | -2.27352 | -4.98263 | Active phosphatase cascade. |

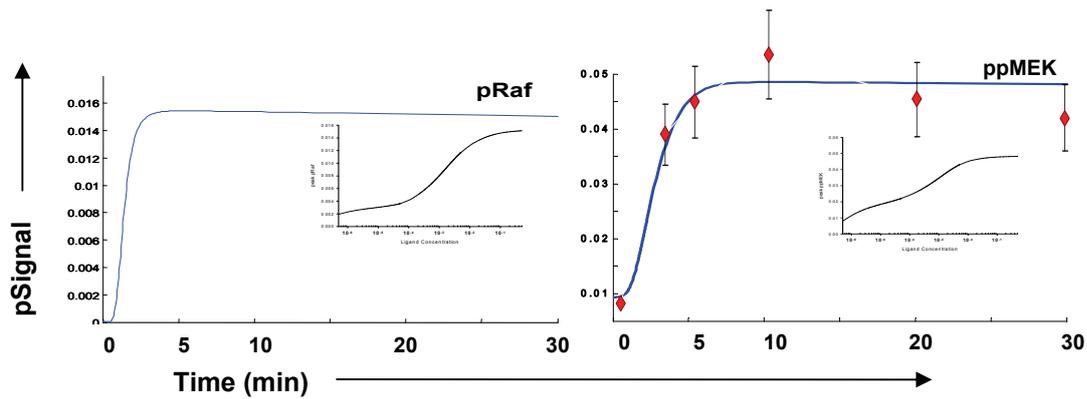

**Figure S2.** Shown here is a simulation of the time course of phosphorylation Raf (pRaf) and MEK (ppMEK) in cells stimulated with a saturating concentration of ligand. The inset in each panel gives the peak level of phosphorylation obtained over a wide range of ligand concentrations. These profiles are consistent with the the experimental results derived in our earlier studies(Kumar et. al., 2007, 2008) The concordance between the experimental (red diamonds, values are mean $\pm$ S.D. of three experiments) and the simulated profiles is shown for ppMEK.

**A**

**B**

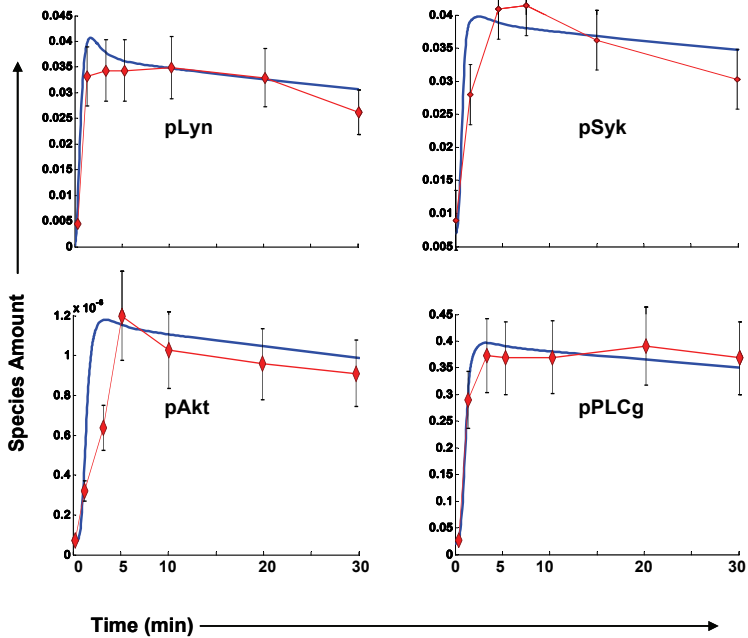

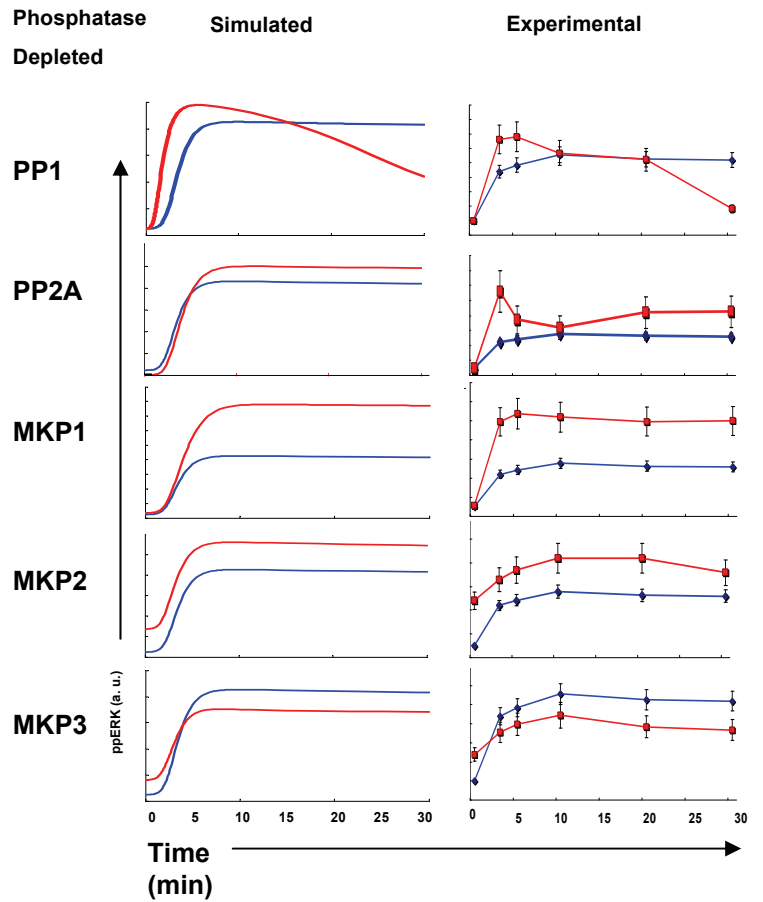

**Figure S3.** Panel A shows simulation of the time course of phosphorylation of the indicated molecules in cells stimulated with a saturating concentration of the ligand. For the purposes of comparison, the experimentally derived values at the various time points are also included (red diamonds, values mean ± S.D. of three experiments). These values are from our earlier study (Kumar et. al., 2007, 2008). For the *in silico* profiles simulations were first run for 3000 seconds to achieve the steady baseline levels of phosphorylations prior to stimulation. This explains the baseline levels of phospho-species seen in these profiles. Panel B compares the experimentally derived (Experimental) and *in silico* (Simulated) profiles of ERK phosphorylation following depletion of cells with the indicated phosphatases. The values for the experimental group were taken from Kumar et. al. (2008). In all cases the X-axes start from zero although the units are arbitrary. In all the panels the profiles obtained in mock (i.e. GFP-siRNA) treated cells are shown in blue whereas those for phosphatase-specific siRNA treated cells are in red. For the *in silico* profiles simulations were first run for 3000 seconds to achieve the steady baseline levels of ERK phosphorylations prior to stimulation. This explains the minor lag period seen in these profiles.

Supplementary Data S4

## A

## Effect of MKP3 knock down by siRNA on MKP1 expression level

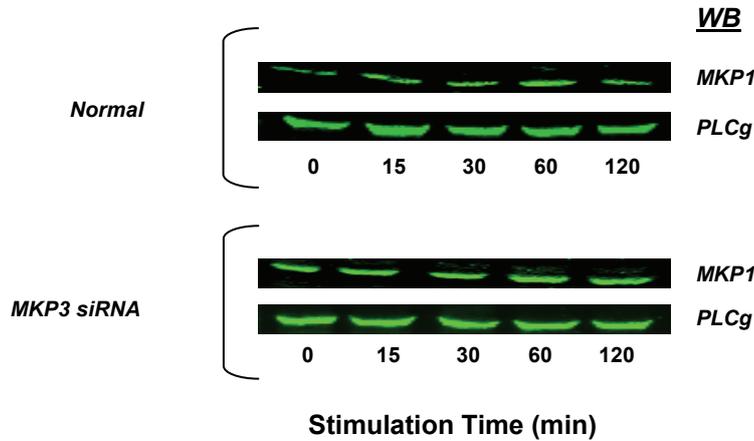

## B

## Activation profiles in the presence of DRB(CK2 inhibitor)

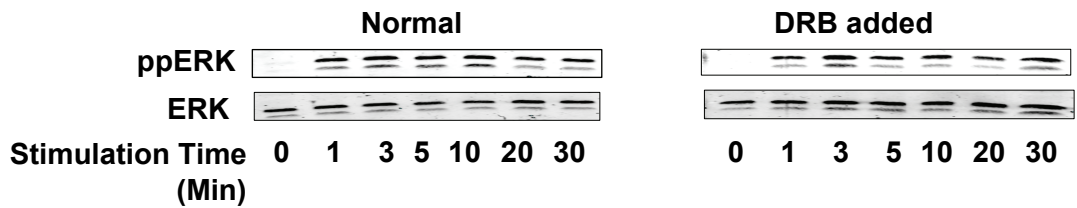

## C

## Graded ERK response for various concentrations of Ligand (Fab2)

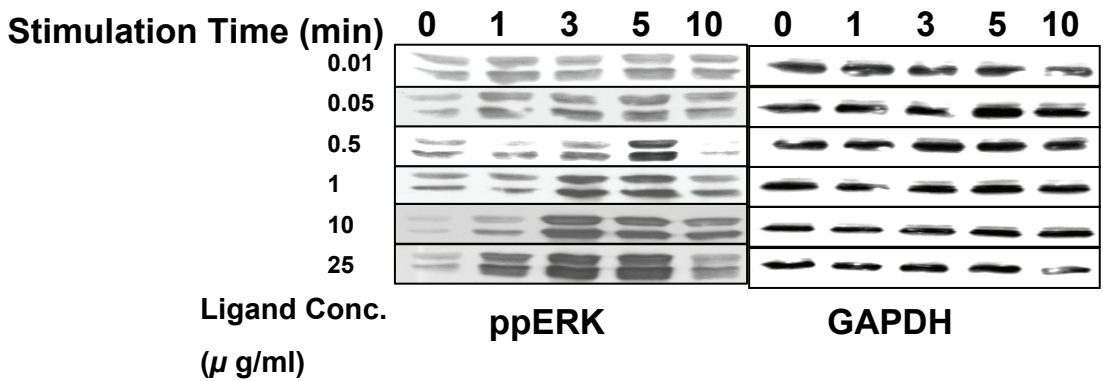

**Figure S4.** **Panel A** shows the Western blot profile of MEK in A20 cells treated either with non-silencing (GFP) siRNA (labeled Normal), or with MKP3-specific siRNA. In both cases the cells were stimulated with the F(ab)$_2$ fragment of anti-mouse IgG for the indicated times. For quantification band intensities were normalized against that obtained for PLCg, which was used as loading control. **Panel B** shows the stimulation time-dependent phosphorylation profiles of ERK (ppERK) in either treated cells, or in cells pre-treated with DRB. GAPDH (Ctrl) served as the loading control for band intensity normalization in these experiments. Shown in **Panel C** are the time dependent phosphorylation profiles of ERK (ppERK) at the indicated concentrations of ligand. Here again, GAPDH was used for the loading control and for normalization of band intensities.

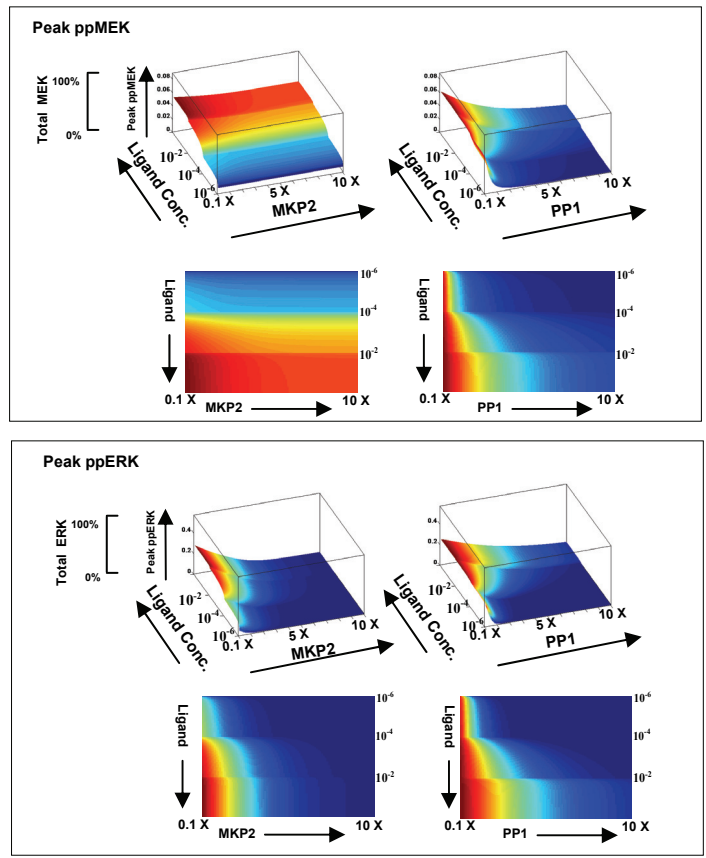

**Figure S5.** The peak phosphorylation of MEK (ppMEK, top subpanel) and ERK (ppERK, bottom subpanel) was determined in response to variations in concentrations of both ligand and either PP1 or MKP2. The remaining details are identical to that described for Figure 5 in the main text.

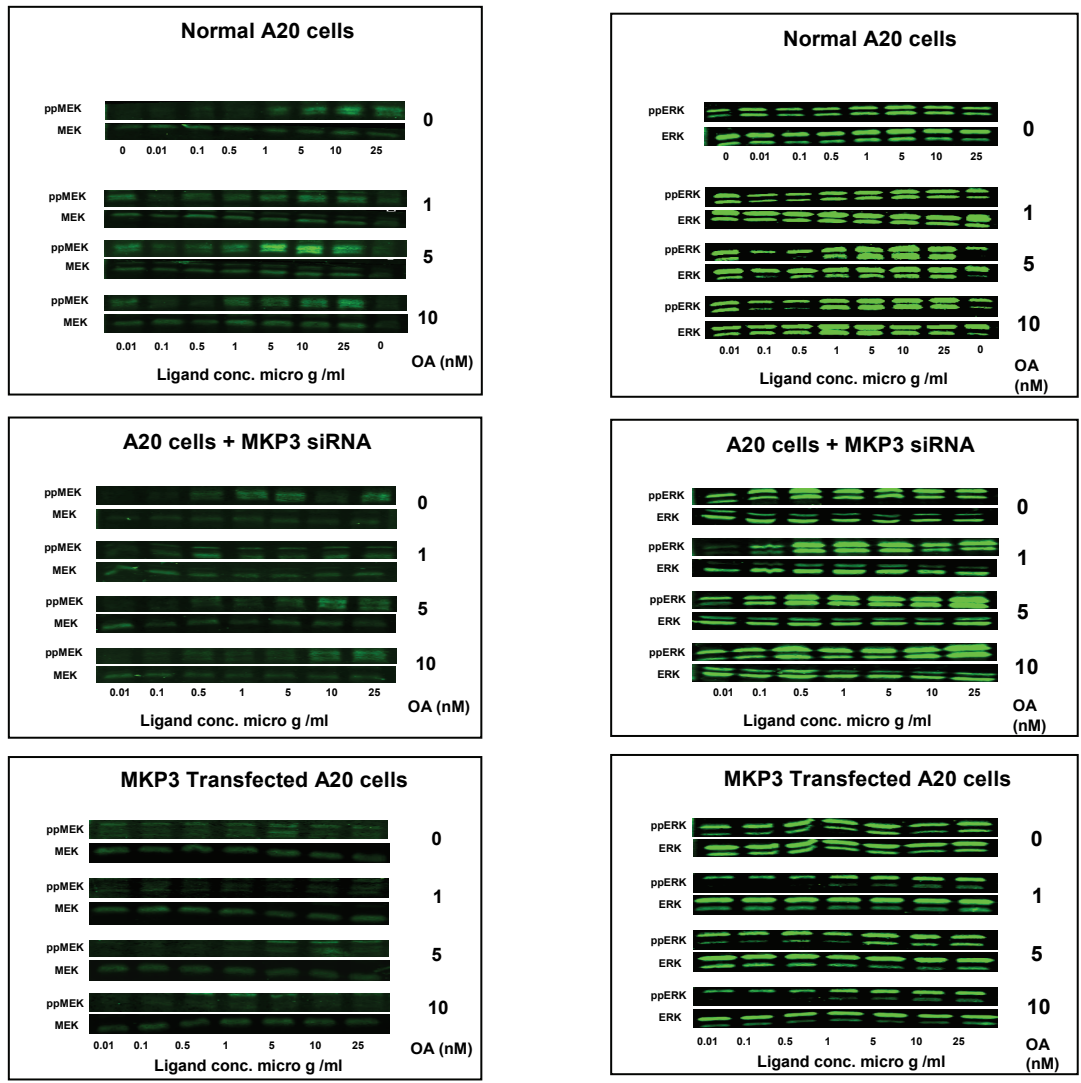

**Figure S6: Ligand dose response of MEK and ERK phosphorylation to variations in levels of phosphatase activity.** Shown here are the Western blot profiles for both the MEK and ERK protein, and its phosphorylated form in either normalA20 cells, in cells treated with MKP3 specific siRNA, or in cells transfected with an MKP3 bearing plasmid (see Experimental Procedures). For each group and condition, cells were stimulated with the indicated concentration of ligand, and the level of phosphorylation monitored 10 min later. The relatively lower intensities of the MEK blots are due to the fact that the concentration of this protein is about 6-fold lower than that of ERK in these cells (Supplementary Data S1)

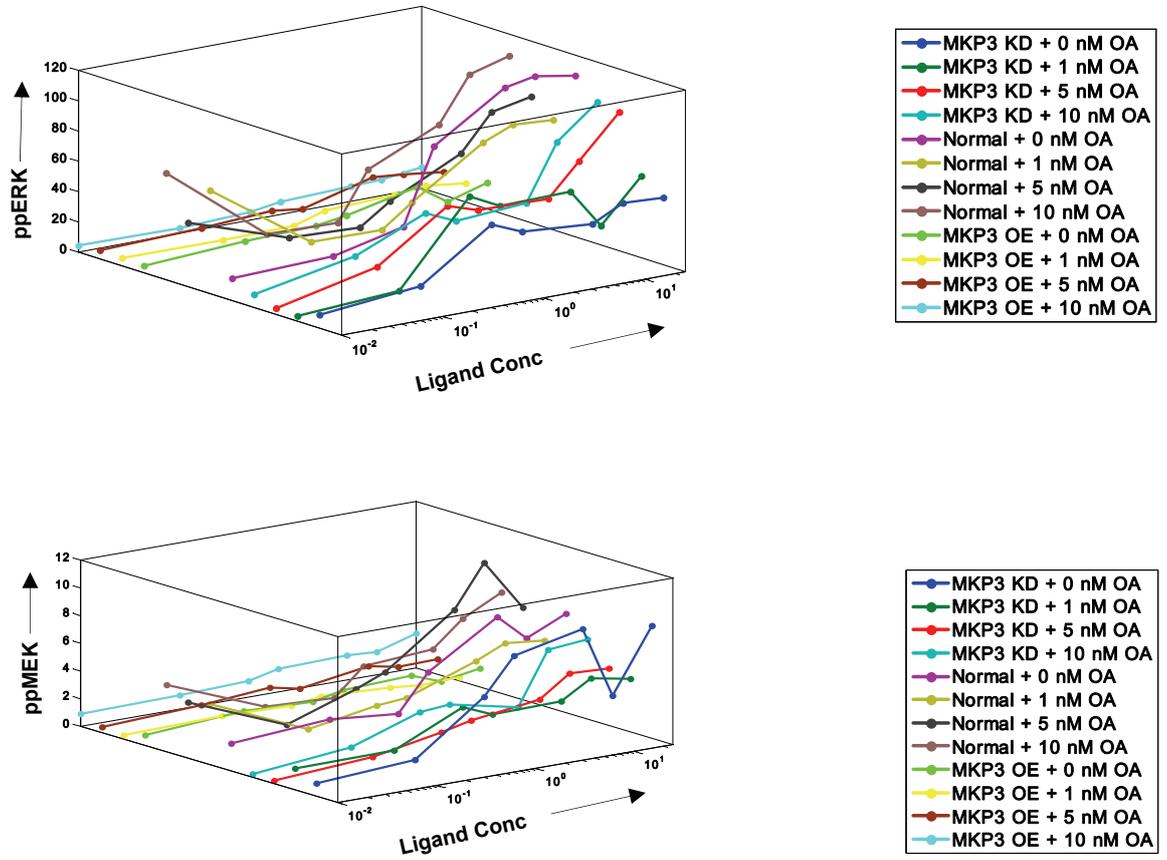

**FigureS7: Modulation of dose response by varying phosphatase concentrations:** Western Blots in Figure S-6 were normalized and quantified using our toolbox (Gel-norm) as described earlier (Kumar et. al., 2007). The plots of these values are shown here.

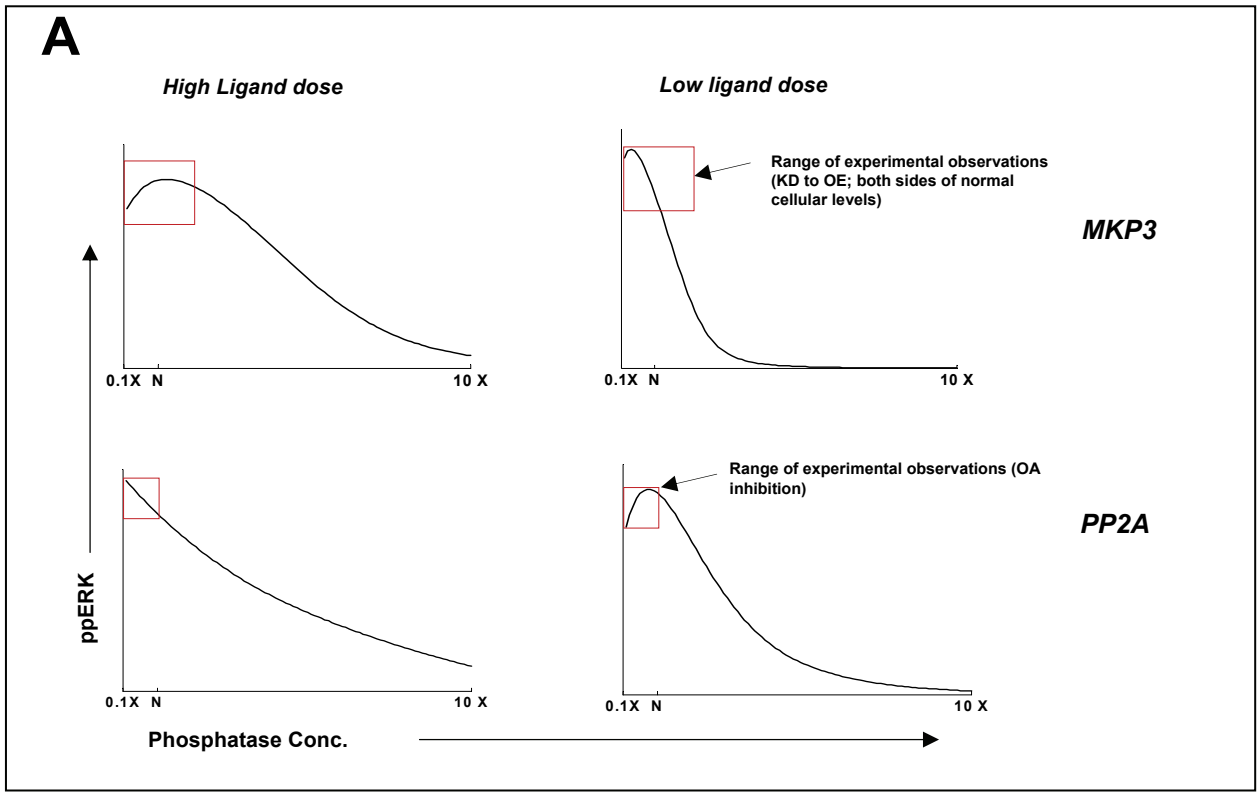

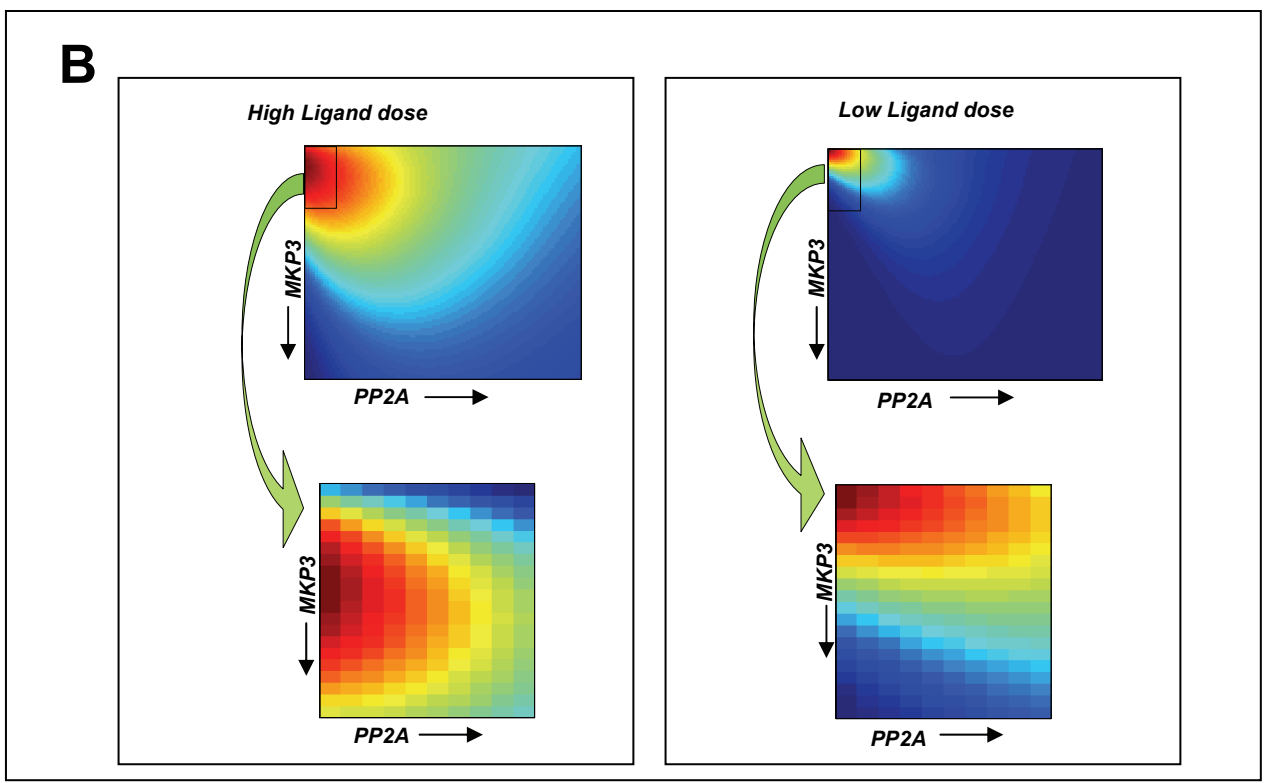

**Figure S8.** Panel A of this figure gives the simulated profile of peak phospho-ERK (ppERK) levels under conditions where concentrations of either MKP3 or PP2A ranged from 10-fold lower, to 10-fold higher than their respective levels in A20 cells. ERK phosphorylation responses at both high and low ligand doses are shown here. In each profile the boxed region indicates the concentration (for MKP3), or activity (for PP2A) range that was taken for the experiments described in Figure 7A of the main manuscript. Similarly MEK profile for prediction were also obtained.

Panel B shows the results of a simulation experiment where peak phospho-ERK responses were monitored under conditions where the concentrations/activities of both phosphatases were derived from the same range as described for Panel A. Here again, results obtained at high and low ligand doses are shown. The top part of this panel gives the results obtained over all the combinations of PP2A and MKP3 concentrations tested, and the boxed region identifies the range represented by the experiments in Figure 7B of the main manuscript. The lower part of this panel shows a magnification of this boxed region.

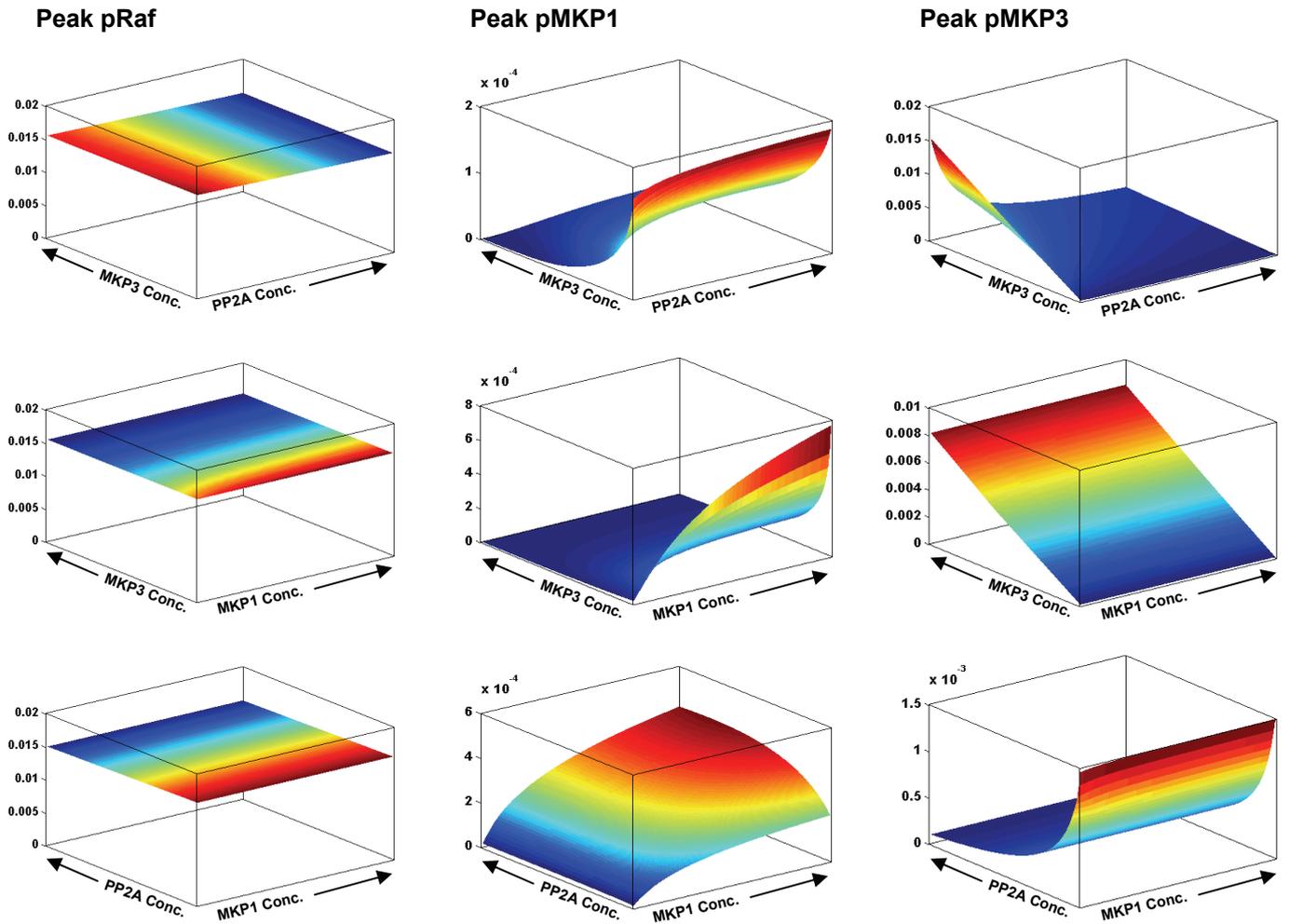

**Figure S9 :** This Figure represents a part of the experiment performed in Figure 6 of the main manuscript. Shown here are the effect of variation in concentrations of the indicated phosphatase pairs on peak phosphorylation levels of Raf (pRaf), MKP1 (pMKP1), and MKP3(pMKP3). It is notable that, as per our expectations from the model in Figure 1A, the amplitude of Raf phosphorylation is only weakly sensitive to changes in concentration of PP2A. According to our model, the effect of altered PP2A levels is buffered by the excess of available PP1.(In all cases arrows showing minimum to maximum conc. of the corresponding enzymes)

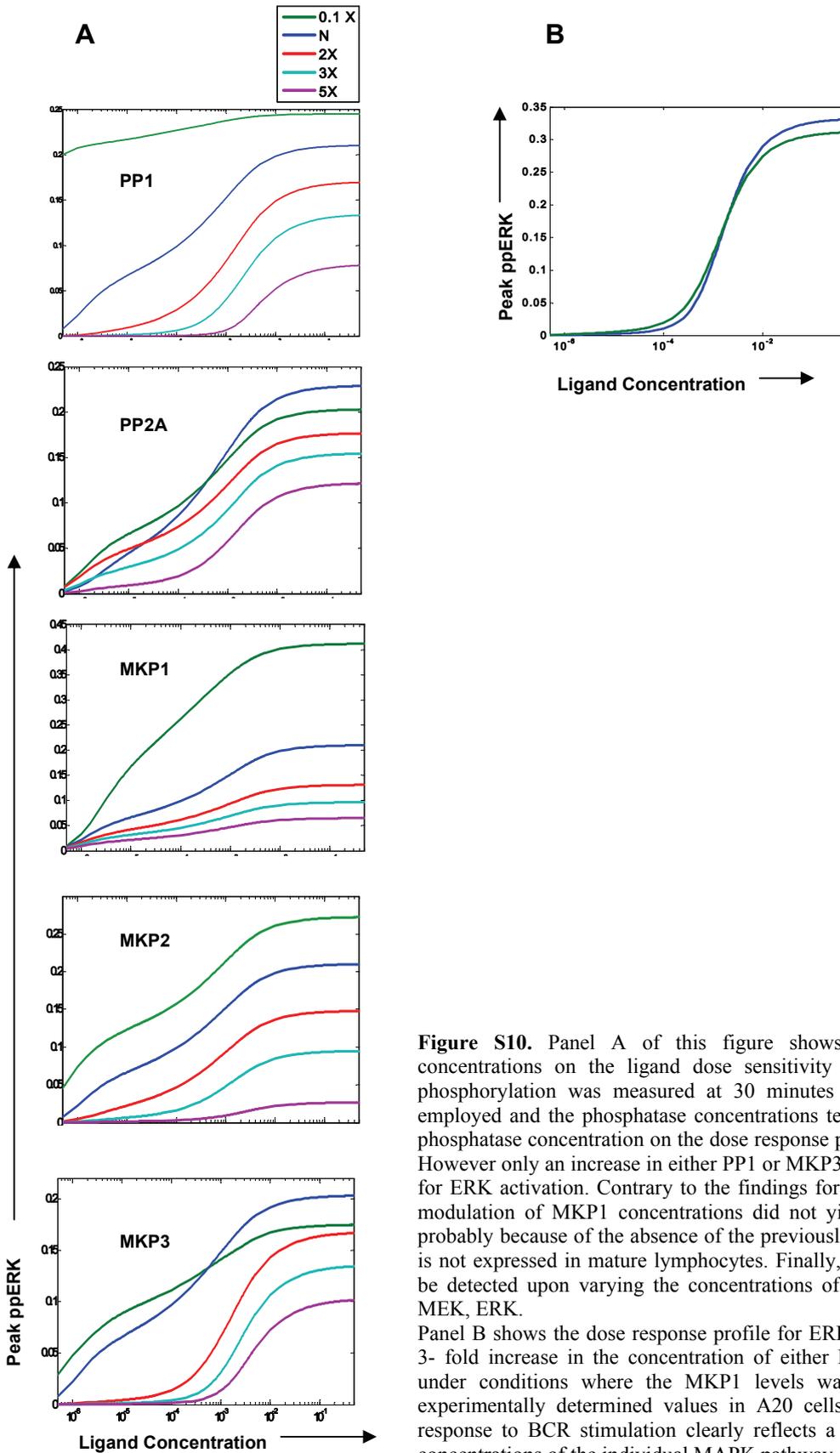

**Figure S10.** Panel A of this figure shows the effect of varying phosphatase concentrations on the ligand dose sensitivity of ERK phosphorylation. Here ERK phosphorylation was measured at 30 minutes of stimulation. Both the ligand dose employed and the phosphatase concentrations tested, are indicated. A marked effect of phosphatase concentration on the dose response profile is clearly evident in all the cases. However only an increase in either PP1 or MKP3 levels led to sharp increase in the slope for ERK activation. Contrary to the findings for another cell line (Bhalla et. al., 2002), modulation of MKP1 concentrations did not yield an ultrasensitive response. This is probably because of the absence of the previously described feedback loop since cPLA2 is not expressed in mature lymphocytes. Finally, no trend towards ultrasensitivity could be detected upon varying the concentrations of the MAPK pathway constituents Raf, MEK, ERK.

Panel B shows the dose response profile for ERK phosphorylation obtained following a 3-fold increase in the concentration of either PP1 (blue line) or MKP3 (Green line) under conditions where the MKP1 levels was reduced by tenfold relative to the experimentally determined values in A20 cells. Thus the shape of the ERK output response to BCR stimulation clearly reflects a combinatorial outcome of the relative concentrations of the individual MAPK pathway-associated phosphatases